\def\ha{{H$\alpha$}}
\def\sii{[\ion{S}{2}]}
\def\LL{\mbox{$\:\lambda\lambda$}}

\documentclass[manuscript]{aastex}
\usepackage{graphicx}
\usepackage{psfig}
\usepackage{epsfig}
\usepackage{color}

\shorttitle{CTIO, ROSAT and Chandra Observations of W28 (G6.4-0.1)}
\shortauthors{Pannuti et al.}

\begin{document}


\title{CTIO, {\em ROSAT} HRI and {\em Chandra} ACIS Observations of the Archetypical 
        Mixed-Morphology Supernova Remnant W28 (G6.4$-$0.1)}


\author{Thomas G. Pannuti\altaffilmark{1}, Jeonghee Rho\altaffilmark{2,3}, Oleg 
Kargaltsev\altaffilmark{4}, Blagoy Rangelov\altaffilmark{4}, Alekzander R. 
Kosakowski\altaffilmark{1,5}, P. Frank Winkler\altaffilmark{6,7}, Jonathan W. 
Keohane\altaffilmark{8}, Jeremy Hare\altaffilmark{4} and Sonny Ernst\altaffilmark{1}}
\altaffiltext{1}{Space Science Center, Department of Earth and Space Sciences, Morehead 
State University, 235 Martindale Drive, Morehead, KY 40351, USA; E-mail: 
t.pannuti@moreheadstate.edu}
\altaffiltext{2}{SETI Institute, 189 Bernardo Avenue, Mountain View, CA 94043, USA; 
E-mail: jrho@seti.org}
\altaffiltext{3}{SOFIA Science Center, NASA Ames Research Center, MS 211-3, Moffett Field, CA 
94035, USA; E-mail: jrho@sofia.usra.edu}
\altaffiltext{4}{Department of Physics, 214 Samson Hall, George Washington University,
Washington, D. C. 20052, USA; E-mail: kargaltsev@gwu.edu}
\altaffiltext{5}{Homer L. Dodge Department of Physics and Astronomy, The University of 
Oklahoma, 440 W. Brooks Street, Norman, OK 73019: E-mail: alekzanderkos@ou.edu} 
\altaffiltext{6}{Department of Physics, Middlebury College, Middlebury, VT 05753, USA; E-mail:
winkler@middlebury.edu} 
\altaffiltext{7}{Visiting Astronomer, Cerro Tololo Inter-American Obervatory. CTIO is operated by
AURA, Inc. under contract to the National Science Foundation.}
\altaffiltext{8}{Department of Physics and Astronomy, Hampden-Sydney College, Hampden-Sydney, 
VA 23943, USA; E-mail: jkeohane@hsc.edu}


\begin{abstract}

We present a joint analysis of optical emission-line and X-ray observations of the archetypical Galactic mixed-morphology supernova remnant (MMSNR) W28 (G6.4$-$0.1). MMSNRs comprise a class of sources whose shell-like radio morphology contrasts with a filled center in X-rays; the origin of these contrasting morphologies remains uncertain. Our CTIO images reveal enhanced [S II] emission relative to H$\alpha$ along the northern and eastern rims of W28. Hydroxyl (OH) masers are detected along these same rims, supporting prior studies suggesting that W28 is interacting with molecular clouds at these locations, as observed for several other MMSNRs. Our {\it ROSAT} HRI mosaic of W28 provides almost complete coverage of the SNR. The X-ray and radio emission is generally anti-correlated, except for the luminous northeastern rim, which is prominent in both bands. Our {\it Chandra} observation sampled the X-ray-luminous central diffuse emission. Spectra extracted from the bright central peak and from nearby annular regions are best fit with two over-ionized recombining plasma models. We also find that while the X-ray emission from the central peak is dominated by swept-up material, that from the surrounding regions shows evidence for oxygen-rich ejecta, suggesting that W28 was produced by a massive progenitor. We also analyze the X-ray properties of two X-ray sources (CXOU J175857.55$-$233400.3 and 3XMM J180058.5$-$232735) projected into the interior of W28 and conclude that neither is a neutron star associated with the SNR. The former is likely to be a foreground cataclysmic variable or a quiescent low-mass X-ray-binary while the latter is likely to be a coronally-active main sequence star.

\end{abstract}




\section{Introduction}

A supernova remnant (SNR) is produced by the interaction between stellar ejecta released
in the violent death of a star in a supernova explosion and the surrounding interstellar medium (ISM). 
SNRs play crucial roles in the evolution of the ISM of a galaxy through such processes as 
cosmic-ray acceleration and the deposition of vast amounts of kinetic energy and 
chemically-enriched material. Our understanding of SNRs has flourished in the modern
era thanks to pointed observations made with space-based observatories such as $\it{Fermi}$,
$\it{Chandra}$ and $\it{Spitzer}$, which are capable of conducting highly sensitive imaging and 
spectroscopic
observations of SNRs in the $\gamma$-ray, X-ray and infrared, respectively. Over 290
Galactic SNRs are known to exist \citep{Green14} based on detections made at wavelengths
across the whole electromagnetic spectrum.
SNRs exhibit a range of morphologies at different wavelengths, which 
define distinct classes into which SNRs have been categorized:
(1) Shell-like SNRs exhibit a shell of emission at both the X-ray and radio wavelengths, 
produced  by an optically-thin plasma and synchrotron
radiation, respectively.  Well known examples include the young remnant Cas A and the more mature Cygnus Loop.
(2) Crab-like SNRs, often known as plerions, feature center-filled morphologies in both X-ray and radio.  At both wavelengths, the observed emission is synchrotron radiation from a nebula surrounding a dominant central pulsar.   (3) A class of composite
SNRs comprises sources that combine characteristics of the previous two classes, typically
exhibiting both an exterior  shell-like structure and a central pulsar-wind nebula.  One well-studied example is CTB 80 \citep{SafiHarb95}. 

A fourth class of 
SNR combines
a shell-like radio morphology with a center-filled centrally-peaked X-ray morphology, but unlike the composite class, 
the central X-ray emission is not non-thermal in origin -- as would be
expected a pulsar-wind nebula -- but is instead
thermal and is produced by an optically-thin plasma. SNRs that belong to this class are known as
``mixed-morphology SNRs" (MMSNRs) \citep{RhoPetre98} or as ``thermal composite" SNRs
\citep{Jones98}. 
MMSNRs typically feature temperature profiles that are relatively uniform, in contrast to  
the radially-dependent  profiles  predicted 
by the classical Sedov solution for SNRs expanding into a uniform ambient medium. Prominent
examples of MMSNRs include W44, CTB 1, HB 3, HB 21 and G352.7$-$0.1 
\citep{Shelton04, Lazendic06, Pannuti10, Pannuti14a}.

\par 
A thorough understanding of the origin of the contrasting morphologies of MMSNRs remains
elusive. In general, MMSNRs appear to be interacting with adjacent
molecular clouds, as indicated by observations of strong infrared line emission from shocked
molecules or OH masers \citep{Frail96, Reach96, Reach98, YusefZadeh03}. 
Therefore, it is
believed that the interaction between the SNR and the molecular cloud dictates the creation
of the contrasting morphologies, though a complete understanding of this phenomenon has
not yet been realized. 
Numerous models have been proposed to explain the origins of the contrasting morphologies
for MMSNRs (for reviews, see \citet{Chen08}, \citet{Vink12} and \citet{Zhang15}). These
models include: (1) enhanced internal gas density due to the evaporation of small and dense 
cloudlets that are overrun by the expanding shock of the SNR \citep{McKee82,White91}; (2)
a hot interior with an elevated density due to a thermal conduction \citep{Cox99,Shelton99};
(3) a radiatively cooled rim and a hot interior \citep{Harrus97,Rho98}; (4) an interaction between the
SNR shock front and the edge of a cloud that appears brighter toward the projected interior
of the SNR \citep{Petruk01,Chen04}; (5) thermal conduction and enhanced emission from
metals due to dust destruction and ejecta enrichment in the SNR interior \citep{Shelton04}; 
and (6) reflection of the SNR blast wave off a wind-blown cavity created by the stellar
progenitor, leading to a powerful reverse shock that  heats the material in the SNR
interior \citep{Chen08}. None of these models can by themselves account for all of the
observed properties of every known MMSNR, and therefore more analyses of these sources
is called for.

\par
Recent spectroscopic X-ray analyses of MMSNRs have revealed the presence of
ejecta-dominated emission as well as evidence that the X-ray-emitting plasmas 
associated with many MMSNRs are over-ionized. By ``ejecta-dominated," we mean 
that the elemental abundances of the X-ray emitting plasma are elevated relative to
solar.  While the presence of elevated abundances is expected for young
SNRs that have not yet evolved past the free-expansion stage, MMSNRs 
have typical ages of between 10$^3$ to 10$^4$ years
and should be well into the Sedov (adiabatic) stage of SNR evolution, dominated by swept-up interstellar matter.
\citet{Zhang15} compiled a list of approximately 30 MMSNRs for which elevated 
elemental abundances have been reported in the literature: this large number of sources
indicates that MMSNRs may have a fundamentally different evolutionary path than
the canonical model of SNR evolution that has been widely accepted in the literature. 
By ``over-ionized," we mean that the cooling rate of these X-ray-emitting plasmas is faster 
than the recombination rate. The ionization state of a plasma
can be characterized by both its electron temperature $T$$_{e}$ and its ionization
temperature $T$$_{z}$, where the latter corresponds to the equilibrium temperature that would give the mix of ions observed spectroscopically.  
When $T_{e} <   T_{z}$, the plasma is said
to be underionized; when $T_{e} >  T_{z}$,  overionized; and when $T_{e} =  T_{z}$, the plasma is said to 
be in collisional ionization equilibrium (see \citet{Miceli11} and \citet{Broersen15} for a more detailed
description of the ionization conditions of the X-ray-emitting plasmas of SNRs). 

\par
The remarkable plasma conditions seen in MMSNRs -- namely the presence of a plasma that appears to be both 
ejecta-dominated and overionized  -- have motivated investigations into how 
such a plasma may develop. \citet{Kawasaki05} presented a scenario that predicts both 
the center-filled thermal morphology of MMSNRs and the overionization of the plasma.
In this scenario, thermal conduction causes the X-ray-emitting plasma to be center-filled:
it also reduces the temperature and density gradients of the hot interior plasma, leading
to a density increase and a temperature decrease: therefore, the plasma becomes
overionized. This intriguing proposed scenario can adequately describe (in general terms)
the observed properties of MMSNRs, but additional investigations are required to
explore it (and its applicability to MMSNRs in general) more fully. 
\par
In this paper, we present 
optical and X-ray observations of the  
archetypical MMSNR W28 (G6.4$-$0.1) made 
with the Cerro Tololo Inter-American Observatory (CTIO) 0.6-meter Curtis Schmidt telescope,
the High Resolution Imager (HRI) aboard {\it ROSAT} and the Advanced CCD Imaging 
System (ACIS) aboard the {\it Chandra} X-ray Observatory. The center-filled 
X-ray morphology of this SNR \citep{Long91, Rho96} -- coupled with its shell-like 
radio morphology \citep{Dubner00} -- are well-known and clearly illustrate the 
morphological properties of MMSNRs. Radio observations
of W28 have revealed a shell-like morphology with luminous northern 
and eastern rims as well as a tangential orientation to the magnetic
field lines over the northern shell \citep{Kundu70, Shaver70, Milne71, 
Kundu72, Frail94a, Dubner00}; a radio spectral index of $\alpha$=0.35$\pm$0.18 
(where $S$$_{\nu}$ $\propto$ $\nu$$^{-\alpha}$) for this source was 
measured by \citet{Dubner00} using VLA data at 328 MHz and 1415 MHz. Narrow-band optical imaging of W28 
\citep{Long91,Mavromatakis04} shows patchy  knots of  emission, combined with 
extended diffuse emission, in both \ha\ and \sii\ lines, especially in the central region where the X-rays are strongest. 
In addition, there are filaments of enhanced (relative to \ha) \sii\ emission -- typical of material heated by SNR shocks -- along the northern and eastern rims, coincident with the brightest portions of the radio shell.  These regions also correspond to the
sites of interaction between W28 and surrounding molecular clouds (see below). 
\par
W28 has been observed with virtually all the pointed X-ray missions, beginning with the {\it Einstein} Observatory, where \citet{Long91} discovered the centrally-peaked thermal X-ray emission from an SNR with a limb-brightened radio shell -- the characteristics that have come to define MMSNRs.  
\citet{Rho02} analyzed {\it ROSAT}
Position Sensitive Proportional Counter (PSPC) and {\it ASCA}
observations of W28 and noted arcs of emission along the northeast and southwest portions 
of the shell, in addition to the brighter centrally peaked emission.
They found that two thermal components
were required ($kT = 0.6$ keV and $kT = 1.8$ keV) to fit the spectrum of 
the central diffuse X-ray emission. This was the first evidence for
thermal variations in the emission from the interior of the SNR and
makes W28 unique among MMSNRs.   They also found that the arc of emission to the southwest
had a hard thermal X-ray component, which they attributed to the breakout of the SNR shock  
opposite to the molecular clouds to the north and east. In their analysis
of a pointed {\it Suzaku} observation made of W28, \citet{Sawada12} fit the extracted 
spectra of the SNR with a multi-ionization-temperature plasma with a common electron
temperature, where the multi-ionization temperatures are interpreted as elemental differences
of ionization and recombination timescales. Such plasma conditions have not been
identified in another MMSNR in the published literature. 

Most recently,  \citet{Zhou14} and \citet{Nakamura14} performed independent 
spectroscopic  analyses of W28 based on {\it XMM-Newton}
observations. \citet{Zhou14} described complex conditions for the plasma at both
the northeastern rim and the bright central region: the spectrum of the northeastern rim
could be fit with either a combination of two thermal components or a thermal
component combined with a power law component, while spectra from the central
region could be fit with either  two thermal components or a single
recombining thermal model. They argued that W28 is expanding into a non-uniform
environment with denser material to the east and north and that the cloudlet evaporation
model may explain the observed properties of the X-ray emission from the center of the SNR
while the thermal conduction model may play a role over large length scales that are
comparable to the radius of the SNR itself.  \citet{Nakamura14} concentrated on the northeastern
region, 
where they showed a bright and twisted portion of the SNR shell to have X-ray emission that could
be fit with a single temperature $kT \approx 0.3$ keV optically thin plasma in ionization equilibrium.  
The emission measure requires a high density, consistent with the interaction with molecular clouds in 
that region.
\par
There is abundant evidence published from other bands that W28 is interacting strongly with 
adjacent molecular clouds to the north and east. As noted previously, 
such a robust interaction with an adjacent molecular cloud seems to be a defining
characteristic of MMSNRs.
Based on observations of broad CO J=1-0 lines with a full-width at half-maximum (FWHM) 
of 11 km sec$^{-1}$ and the presence of warm dense clouds, \citet{Wooten81}
first suggested that W28 is interacting with adjacent molecular clouds. This result was confirmed 
through subsequent CO observations of the J=3-2 and J=1-0 lines that identified the 
presence of shock-excited masers along the northern and eastern rims of the SNR
\citep{Frail94b, Frail98, Arikawa99, Claussen97, Claussen99, Hoffman05, Hewitt08}.
Mid-infrared observations  of W28 from space-based observatories have also confirmed
that W28 is interacting with nearby molecular clouds: using observations from the 
{\it Infrared Space Observatory}, \citet{Reach00} reported the detection of infrared S(3)
and S(9) rotational lines of H$_2$, along with strong emission in atomic fine-structure mid-IR lines, 
especially  [\ion{O}{1}] $\lambda$\,64$\mu$m.  Spectral fits indicate this emission
originates from regions of density $\sim 10^3\;{\rm cm}^{-3}$.    
In addition, \citet{Neufeld07} and \citet{Marquez-Lugo10} presented 
spectroscopic and imaging observations, respectively, made of W28 with the {\it Spitzer} Space 
Telescope. The former paper described the spectroscopic detection of both a shock-excited
component and a low-density diffuse emission component for the H$_2$ S(0) line, while 
the latter paper presented an analysis of the spectral energy distributions of Class I young
stellar objects \citep{Lada87} seen in projection toward the northwestern interaction region of W28 and 
suggested that the formation of these sources may have been triggered by the supernova 
event associated with this SNR. 
\par
Finally, high-energy $\gamma$-ray emission has been detected form W28 by {\it Fermi}
\citep{Abdo10} and H.E.S.S. \citep{Aharonian08}: modeling of this emission by those authors 
and \citet{Li10} favor a hadronic origin where energetic protons collide with molecular clouds, 
again supporting the scenario that W28 is indeed interacting with adjacent molecular clouds. 
The totality of evidence
that the SNR shock in W28 is interacting with one or more molecular clouds in the north and east
is compelling.  
In this paper, we will consider high angular
resolution optical and X-ray observations of W28 with the goal of analyzing its
high energy emission on fine spatial scales and testing the applicability of
the different models in explaining its X-ray properties.  
\par
As is the case for many Galactic SNRs, the distance to W28 is poorly 
known, with estimates ranging between 1.6 and 3 kpc 
\citep{Clark76, Milne79, Frail94a}. For the present 
paper, we note that CO observations of the shock interaction between W28 and 
an adjacent molecular cloud imply a distance of 1.8 kpc \citep{Arikawa99} based on the 
rotation curve for a Sun-Galactocentric distance of 8.5 kpc \citep{Clemens85}. We therefore
adopt this distance to W28; we note that this choice of distance is consistent with 
our previous X-ray study of W28 \citep{Rho02}. 
\par
The present paper is organized as follows: in Section 
\ref{ObservationSection} we describe the observations that were
incorporated into our study, beginning with the optical observations made with 
the CTIO 0.6-meter Curtis Schmidt telescope
(Section \ref{OpticalObsSection}) and continuing with high angular resolution X-ray observations
made with the {\it ROSAT} High Resolution Imager (HRI)
(Section \ref{HRIObsSection}) and the {\it Chandra} Advanced CCD for Imaging and
Spectroscopy (ACIS) (Section \ref{ACISObsSection}). 
Our imaging and spectral
analyses of the datasets are presented in Section \ref{AnalysisSection}: we discuss our
narrowband optical images in Section \ref{OpticalImagesSubsection}, our 
{\it ROSAT} HRI mosaic of W28 in Section \ref{ROSATHRIMosaicSubsection}, our spatially 
resolved X-ray spectroscopic analysis using {\it Chandra} in Section 
\ref{ChandraSpectralSubsection}, and the morphologies of W28 at different wavelengths in 
Section \ref{SpatialComparisonSubsection}. In Section \ref{DiscreteXraySourceSection} we 
discuss X-ray emission from two particular discrete X-ray sources seen toward W28, namely
CXOU J175857.55$-$233400.3 (a hard X-ray source seen in prior {\it ASCA} observations
of W28 as reported by \citet{Rho02}) and 2XMM J180058.6$-$232724 = 3XMM 
J180058.5$-$232735 (a discrete X-ray source seen toward the geometric center of W28).
The analyses of these sources is conducted to determine if either source may be an
X-ray-emitting neutron star that is physically associated with W28. Finally, the conclusions
of this paper are presented in Section \ref{ConclusionsSection}.

\section{Observations\label{ObservationSection}}

In this Section, we present a brief description of the observations
(as well as the corresponding data reduction) for the CTIO, {\it ROSAT} HRI and
{\it Chandra} ACIS observations of W28. In addition, in our analysis
we also make use of a 1415 MHz radio map of W28 based on VLA data: this map was kindly 
provided to us by G. Dubner, and details about this map can be found in \citet{Dubner00}.  

\subsection{Optical Observations\label{OpticalObsSection}}

We obtained optical images of W28 from the 
CTIO 0.6-m f/3.5 Curtis Schmidt telescope on the night of 1998 June 22 (UT).  The telescope was 
equipped with the SITe2K\#5  CCD, mounted at Newtonian focus, to give a field 1\fdg32 square at a 
scale of $2\farcs 32\; {\rm pixel}^{-1}$.
We used narrow-band interference filters centered on  the \ha\ and \sii\ emission lines, plus a matched 
line-free continuum filter for subtracting the stars to better reveal faint nebulous  emission.  The 
observational details are given in Table \ref{OpticalObsTable}.  
The data were processed using conventional IRAF\footnote{IRAF is distributed by the National 
Optical Astronomy  
Observatories, which is operated by the  AURA, Inc. under cooperative 
agreement with the National Science Foundation.} techniques, including bias subtraction and 
flat-fielding using a series of well-exposed twilight sky flats.  A World Coordinate System, based 
on stars from the UCAC1 catalog \citep{Zacharias00}, was placed on each individual frame; all the 
frames were then transformed to a common system at a finer scale of 1\arcsec\,pixel$^{-1}$ and 
stacked by filter. The continuum image was scaled 
appropriately and subtracted from the \ha\ and \sii\ ones to effectively remove most of the stars.  The 
final images were flux-calibrated based on observations of several spectrophotometric standard stars 
from the list of \citet{Hamuy92}.

\subsection{{\em ROSAT} HRI Observations\label{HRIObsSection}}

W28 was the subject of nine pointed {\it ROSAT} HRI observations conducted between 20 March 1995
and 11 September 1998 that
sampled the entire X-ray extent of the SNR. The HRI was sensitive to 
photons over the energy range between 0.1--2.4 keV and had a field of view 36\arcmin\ square 
\citep[see][for details]{Zombeck95}.
With an angular size of $\sim 48\arcmin$ \citep{Green14}, W28 
is larger than the HRI field; hence we generated a mosaicked image of W28 
with data from the individual exposure-corrected HRI observations. 
We used the techniques described by \citet{Snowden98} for analyzing 
{\it ROSAT}~HRI images and the Extended Source Analysis Software
(ESAS\footnote{Obtained from 
$\tt{ftp://legacy.gsfc.nasa.gov/rosat/software/fortran/sxrb/.}$}) which was
created by S.L. Snowden and K.D. Kuntz of Goddard Space Flight Center.
Images were prepared from the nine individual observations by first
casting the observed events, the exposure maps and the
background maps into images, and then using these images to generate 
intensity images for each individual observation. Six of the nine
observations were ``broken exposures" which sampled the same region of W28
more than once; that is, the central, southern and southeastern portions
of W28 were each observed twice (see Table \ref{Table1} for details). Before 
making the mosaic image, we summed 
each pair of broken exposures to make total intensity images, exposure maps
and background maps for these three portions of W28. We then mapped
the counts, exposures and background counts from the projection of the
individual exposures into the mosaic projection. We next calculated
the Long-Term Enhancement -- that is, the intrinsic HRI background 
\citep{Snowden98} -- for each exposure (or pair of broken exposures):
this step determines the relative offsets in the zero levels of the 
observations to be merged. We made a mosaicked image of the modeled 
offset counts and then finally a mosaicked final count rate image (with
a pixel size of 12\arcsec) using the individual
counts, the modeled background counts, the modeled offset counts
and the exposure maps for each individual exposure (or pair of broken
exposures). We adaptively smoothed the full mosaicked rate map, using a 
smoothing kernel of 30 counts for the adaptive filtering algorithm. 

\subsection{{\em Chandra} ACIS Observations\label{ACISObsSection}}

W28 was observed with the ACIS detectors on the {\it Chandra} X-ray Observatory on
12 October 2002 (ObsID 2828; PI Rho).  The pointing placed the back-illuminated S3 chip (which has 
enhanced soft X-ray sensitivity) on the brightest portion of the central emission; the adjacent S2 and 
S4 chips sampled other portions of the central emission, while two ACIS-I chips sampled emission in 
the southwest of W28, where the hardest emission is located \citep{Rho02}. 
Data were processed with CIAO version 4.8 (CALDB Version 4.7.2), beginning with the CIAO 
tool $\tt{chandra\_repro}$ to assure that the latest calibrations were applied.  We further filtered the 
resulting events file to exclude periods with significant background flare activity, resulting in a net 
effective exposure time of 
88994 seconds. In Table \ref{Table1} we present a summary of the {\it Chandra}
observation of W28. 
\par
To perform a spectral analysis of the X-ray-emitting plasma associated with W28, we divided
the field of view of the ACIS-S3 chip into multiple regions and used the CIAO tool
$\tt{specextract}$ to generate the files -- source spectra, background spectra, ancillary
response files (ARFs) and redistribution matrix files (RMFs) -- that are necessary for spectral
analysis. We considered diffuse emission from W28 as sampled by the ACIS-S3 and -S4 chips
during these observations: no diffuse emission was detected at a significant level from W28 by the ACIS-S1
chip during this observation and we do not consider that chip in the present paper.
Because the entire S3 chip was filled by diffuse emission from W28, a blank sky observation
was used to extract background spectra: the pointer header keyword values from the 
observation of W28 were applied to the blank sky observation file. The blank sky observation
file was also then reprojected to match the projection of the ACIS chip: the same region used
for source spectrum extraction in the W28 dataset was used in the reprojected blank sky dataset
to obtain background spectra. The extracted spectra were grouped to a minimum of 25 counts
per channel. 
To reduce the effect of point source contamination on the extracted spectra (from background sources or Galactic stellar sources seen in projection
toward W28), the CIAO tool $\tt{wavdetect}$ 
was used to identify unresolved sources detected by the {\it Chandra}
observation. Fluxes from the positions of these identified point sources were excluded when
analyzing the extracted diffuse emission. Analysis of the extracted spectra was conducted using
the XSPEC software package \citep{Arnaud96} Version 12.9.0n.

\section{Analysis and Results\label{AnalysisSection}}

\subsection{Narrow-band Optical Images of W28\label{OpticalImagesSubsection}}

In Figure~{\ref{figure_ha_sii_pair} we show the optical images of W28 in the \ha\ and \sii\ emission lines, after continuum subtraction to remove most of the stars in this highly congested field and to 
better reveal faint nebulous emission.   The images in both lines show the same patchy diffuse emission in the central portion of W28  seen in previous optical images \citep{Long91, Mavromatakis04}. In addition, there is more filamentary emission, especially near the northern and eastern rims of the shell, that is stronger in \sii\ relative to \ha\ -- as is typical of SNR shocks.
The differences between the two bands are more apparent in the two color Fig.~\ref{figure_w28_color}, where \ha\ is shown in red and \sii\ in green.  The identical image is also shown in Fig.~\ref{figure_w28_color_overlay}, overlaid with radio contours from \citet{Dubner00}.  
The locations of OH masers --  signposts
of interaction between an SNR and a molecular cloud -- as detected by \citet{Claussen99} are
indicated with magenta crosses: notice how the locations of the masers more strongly
correlate with the locations of high \sii\ emission. Such a correlation may be expected, given that
elevated \sii\ emission relative to H$\alpha$  is a known tracer of SNR shocks.  There is a particularly dense concentration of masers seen toward
the northeastern rim (also known as an ``ear"-like structure of emission -- see \citet{Rho02}):
this rim -- which is also detected in the X-ray and radio -- is also detected prominently in the \sii\ image. 

\subsection{{\em ROSAT} HRI Mosaic\label{ROSATHRIMosaicSubsection}}

To help provide a context about the observed morphologies of W28 at different 
wavelengths, in Figures \ref{W28VLARadioMasers} and \ref{W28ROSATPSPC}
we present radio and X-ray images, respectively, of the SNR. The shell-like radio 
morphology of the SNR is apparent, and the SNR appears to be brightest along its northern
and eastern rims, where the SNR is interacting
most dramatically with adjacent molecular clouds, as indicated by the positions of the OH masers.
The X-ray map was made using the {\it ROSAT} PSPC and was published by \citet{Rho02}. 
Inspection of these two figures helps to illustrate the stark contrast in morphologies at the 
two wavelengths. The X-ray emission from W28 appears to be centrally concentrated at 
approximately RA (J2000.0) 18$^h$ 00$^m$ 26.$^s$5 and Decl. (J2000.0) 
$-$23$^{\circ}$ 24$\arcmin$ 24$\arcsec$
with faint diffuse extensions filling much of the interior volume of W28. The only feature
seen in both the X-ray and the radio is the bright northeastern rim which is also
prominent in the \sii\  image and the site of a high concentration of
masers, as described in the previous section.

\par
In Figure \ref{W28ROSATHRIMasers} we present our {\it ROSAT} HRI mosaic image of W28, which
covers the entire angular extent of the SNR. Like the
{\it ROSAT} PSPC image  in Figure \ref{W28ROSATPSPC},
this image demonstrates the centrally-concentrated nature of the X-ray emission from the
SNR. With its angular resolution superior to that of the PSPC, the mosaicked HRI
image shows more clearly the difference between emission in the X-ray and radio bands.
\citet{Zhou14} presented a mosaicked {\it XMM-Newton} image of W28 with comparable
angular resolution to this mosaicked image and describe the X-ray morphology of the SNR
in their image in similar terms. 

\subsection{{\em Chandra} Imaging and Spectra of Individual Regions
\label{ChandraSpectralSubsection}}

In Figure \ref{W28Chandra3Color} we present an exposure-corrected, adaptively 
smoothed, three-color {\it Chandra} image of W28: the red, green and blue emission correspond to 
soft, medium and hard X-ray emission which in turn correspond to the energy ranges of 0.5 keV - 1.2 
keV, 1.2 keV - 2.0 keV and 2.0 keV - 7.0 keV, respectively. 
As indicated by previous X-ray observations of W28, the emission is centrally concentrated:
the ``yellow" nature of the emission detected from the central region indicates that this emission
is not entirely soft (which is typical for SNRs) but that harder emission is present in this region
as well.  A clear decreasing gradient in brightness is seen in the direction toward the radio rim
of the SNR: the centrally-peaked emission and decreasing brightness gradient are typical of
MMSNRs like W28. 
Toward the southwestern portion of the SNR, the hard emission detected and described
by \citet{Rho02} appears to be concentrated into a single discrete hard X-ray source. The true
angular extent of this source is hard to determine given the elongation of the point spread function
of {\it Chandra} for an object located so far from the optical axis of the telescope. A more detailed discussion of this source is provided in Section \ref{DiscreteXraySourceSection}.
\par
We conducted a spatially-resolved spectroscopic analysis of the X-ray emission form 
W28 in three different steps: we extracted and fitted spectra for the bright central peak
that was located on the ACIS-S3 chip during the observation,
for five individual regions located on the ACIS-S3 chip and one region on the ACIS-S4
chip (this latter region appears to be associated with both strong radio and [\ion{S}{2}] emission
along with pronounced ``yellow" X-ray emission) and for eight annular regions located on the
ACIS-S3 chip that are centered on the bright central X-ray peak along with the region on the
ACIS-S4 chip. Below we present
and discuss the results of these fits below in turn. 
\subsubsection{Spectrum of Bright Central Peak}
First, we extracted a spectrum from the central peak of
emission and used XSPEC to fit the extracted spectra with different thermal models: (1) two thermal 
plasmas in collisional ionization equilibrium with different temperatures
and with variable elemental abundances (VAPEC+VAPEC)\footnote{This model is based
on a variable elemental astrophysical plasma emission code (APEC). For more information
about this code, see \citet{Smith01} and $\tt{http://www.atomdb.org}$.}; (2)  two thermal plasmas in 
non-equilibrium ionization with different temperatures and with variable elemental abundances 
(VNEI+VNEI); (3) a more sophisticated model of a single thermal recombining plasma with variable 
elemental abundances, where the plasma is assumed to have started in collisional equilibrium with the 
initial temperature $k$$T$$_{\rm{Z}}$ and currently has an electron temperature $k$$T$$_{\rm{e}}$ 
(VRNEI); and finally (4) two thermal recombining plasmas with variable elemental abundances and 
different values of $k$$T$$_{\rm{Z}}$ and $k$$T$$_{\rm{e}}$ (VRNEI+VRNEI).
The reader is referred to \citet{Borkowski01} for a thorough description about non-equilibrium 
ionization models. 
Note that non-equilibrium models are characterized by the ionization timescale parameter $\tau$: this 
parameter -- which is the product of the electron number density $n$$_{\rm{e}}$ and a timescale $t$ 
-- indicates whether or not the X-ray emitting plasma is in collisional ionization equilibrium. 
In the cases where two thermal models were used simultaneously (namely the VAPEC+VAPEC
fit, the VNEI +VNEI fit and the VRNEI+VRNEI fit), the elemental abundances of the two
models were tied together.
All of these thermal models were multiplied by the T\"{ubingen}-Boulder interstellar medium
absorption model TBABS to account for absorption along the lines of sight. For elemental abundances, 
we adopted the set from \citet{Wilms00}.
\par
The region of spectral extraction for the bright central peak is shown in Figure
\ref{W28BrightCentralPeakFigure}.
In Table \ref{BlobSpectrumTable} we present the results of our fitting: we find that only the 
TBABS$\times$(VRNEI+VRNEI) model provides a statistically-acceptable fit to the extracted
spectrum.
The fit with the TBABS$\times$(VRNEI+VRNEI)
model yields two low-temperature thermal components ($k$$T$$_{\rm{e1}}$ = 0.10 keV and $k$$T$$_{\rm{e2}}$ =
0.42 keV): the corresponding $k$$T$$_{\rm{Z}}$ values for these thermal plasmas are 0.16 keV and 3.08 keV,
respectively. Neither of these thermal plasmas are in collisional ionization equilibrium ($\tau$$_{\rm{1}}$ and 
$\tau$$_{\rm{2}}$ are both significantly less than 10$^{12}$ cm$^{-3}$ s). During the fitting
processes, the abundances of different elements were allowed to vary: we found that only
the abundances of oxygen, neon, magnesium and iron showed evidence for being either
underabundant or overabundant relative to solar. When the abundances of magnesium,
silicon, sulfur, argon and calcium were allowed to vary, we found no evidence that the abundances
of any of these elements were significantly different from solar: therefore, we left the abundances
of these elements frozen to unity during the fitting process. The fit with the 
TBABS$\times$(VRNEI+VRNEI) model exhibits elemental abundances of oxygen, neon and
iron to all be subsolar: this result indicates that the X-ray emission from the bright central peak is
dominated swept-up ISM rather than stellar ejecta. We present fits to the extracted spectrum of this bright central region
using these four models in Figure \ref{CentralRegionSpectralFigure}. Our results are broadly consistent with other 
published spatially-resolved X-ray spectroscopic analysis of W28: for example, \citet{Sawada12} used multiple
VRNEI components to obtain a statistically-acceptable fit to the extracted {\it Suzaku} XIS spectra of the SNR. 
\subsubsection{Spectra of Five Regions on the ACIS-S3 Chip and A Region on the ACIS-S4 Chip}

The second step of our spatially-resolved spectroscopic analysis of the X-ray emission from W28 was
to extract and fit spectra from five regions seen toward the central bright emission of the SNR 
on the ACIS-S3 chip and a region on the ACIS-S4 chip. These regions are shown in Figure 
\ref{ChandraSpecExtractionFigure}: we refer to these as Regions 1 through 5 along with the ACIS-S4
chip region for the remainder of this paper. The spectra from the five regions on the ACIS-S3 chip 
were fitted over the energy range from 0.5 to 4.0 keV while the spectrum from the region on the 
ACIS-S4 chip was fitted over the energy range from 0.6 to 2.0 keV due to poorer statistics in
this spectrum. These spectra were fitted with three different thermal
models:  VNEI,  VRNEI, and VRNEI+VRNEI. We first compare the fits made with the VNEI and
the VRNEI models. In fitting the spectra with these two models, the column density
for all of the regions was tied together and treated as a single parameter: we found that fitting
the column densities individually for the five regions produced no statistically significant 
differences in the fits. In all the fits, the elemental 
abundances of oxygen and iron were 
allowed to vary, while the elemental abundances of the other elements were frozen to solar values.
The results of these spectral fits using the VNEI and VRNEI models 
are presented in Tables \ref{GroupVNEITable} and
\ref{GroupFitVRNEITable}. We find that the
VRNEI model provides a superior fit to the spectra than the VNEI model:
we interpret this result to indicate that the X-ray emitting
plasma associated with W28 is best described on small spatial scales by an overionized 
recombining plasma. 
\par
Finally, in Table \ref{SixRegionsVRNEIVRNEIFitsTable}, we present the results
of fitting the spectra with a TBABS$\times$(VRNEI+VRNEI) model where again the column density
for all of the regions was tied together and treated as a single parameter. We have also allowed 
the abundances of
neon to vary as well as oxygen and iron. We find that -- when comparing the values of 
$\Delta$$\chi$$^2$ for these fits to the fits presented in the previous tables -- the VRNEI+VRNEI
model produces a greatly improved fit result with a dramatically reduced value of 
$\Delta$$\chi$$^2$ down to 1.24. In Figure
\ref{SixRegionsSpectraFigure}, we show the spectra of these six regions as fit with the 
TBABS$\times$(VRNEI+VRNEI) models and the contributions of the two individual thermal 
components in each fit are shown. 
\par
We also find evidence of enhanced elemental abundances
of oxygen relative to iron in the extracted {\it Chandra} spectra of W28: this result lends support
to the scenario that W28 was produced by a massive stellar progenitor 
\citep{Woosley95,Nomoto97}. In 
previous work, we have used the ratio of oxygen to iron elemental abundances as seen in the 
extracted {\it Chandra} spectra of the MMSNR CTB 1 (G116.9$+$0.2) to estimate the progenitor mass as 10 and 15 
M$_{\odot}$ -- see \citet{Pannuti10}). By inspection of the parameters of the fits presented
in Tables \ref{GroupVNEITable} through \ref{SixRegionsVRNEIVRNEIFitsTable}, it appears that the ratios of
the elemental abundances of oxygen to iron (relative to solar abundances) range up to approximately 
3: this result is broadly similar with the results of 
our spectral fitting for CTB 1 and suggests that the progenitor had approximately the same mass.
This result lends support to the scenario that W28 had a massive stellar progenitor and that
-- as a class -- MMSNRs tend to stem from massive stellar progenitors.
\par
Our spectral fitting results are again consistent with the analysis of extracted {\it Suzaku}
spectra of W28 as presented by \citet{Sawada12}: the superior angular resolution of {\it Chandra}
allows us to probe the properties of the plasma on much smaller spatial scales. 

\subsubsection{Spectra of Annular Regions on the ACIS-S4 Chip and A Region on the ACIS-S4 Chip}

Lastly, we discuss our analysis of spectra extracted from annular regions centered on the
bright central peak of X-ray emission on the ACIS-S3 chip along with the region on the
ACIS-S4 chip. As described elsewhere, MMSNRs are known 
to feature uniform temperatures in their interiors which is believed to
result from thermal conduction \citep{Shelton04,Vink12}. Furthermore, detailed magnetohydronamic 
simulations of the evolution of MMSNRs predict that thermal conduction in the interiors of these
sources will facilitate a steep metallicity gradient in the extracted X-ray spectra of the SNR
\citep{Orlando09}. To search for variations in temperatures and metallicities as a function of
radius in the central thermal X-ray emission, we extracted and fit spectra from eight annular
regions (each with a thickness of approximately 25 arcseconds) that centered on the bright
X-ray peak located on the ACIS-S3 chip. The annular regions of spectral extraction are shown
in Figure \ref{AnnularRegionsFigure}: note that these regions extend nearly 
to the edge of the chip itself.
\par
The spectra of the annular regions were fit simultaneous with the spectrum from the 
ACIS-S4 region with a TBABS$\times$VRNEI model. The abundances
of oxygen and iron were allowed to vary while the abundances of the other elements were 
frozen at solar. The results of these fits are presented in Table \ref{AnnularRegionsTable}. We
find that the values for electron temperature $T$$_{\rm{e}}$ remain approximately constant
for all of the regions (as seen in other MMSNRs): however, as indicated by the fitted abundances
of oxygen and iron for all of the regions, we find no evidence for a gradient in metallicity in
the X-ray-emitting spectra. In predicting the presence of such a gradient, \citet{Orlando09}
argue that such a gradient would be due to a reverse shock propagating inward and confining
the ejecta to the inner part of the SNR. It is possible that in the evolution of W28, a reverse
shock propagating inward has not established this gradient: the absence of the gradient may
be due to the complex ambient ISM into which W28 is expanding. Additional observations and
analysis are needed to explore this result further.

\subsection{Spatial Comparison of X-ray, Optical and Radio Emission 
from W28\label{SpatialComparisonSubsection}}

We have presented a comparison of the spatial morphologies of W28 at X-ray, optical and
radio wavelengths. As the archetypical MMSNR, W28 features a shell-like radio morphology
with a contrasting center-filled X-ray morphology: this X-ray morphology is defined by thermal
emission. The only spatial feature that is detected at all three wavelengths is the northeastern
``ear," where the W28 shock appears to be interacting strongly with 
an adjacent molecular cloud, and where there is  a high density  of OH
masers \citep{Claussen97}. The detection of emission at multiple
wavelengths from particular features of an MMSNR underscores that such
features may be sites of particularly vigorous interactions between the SNR and the
adjacent molecular clouds.
\par
We note that the optical morphology of W28 exhibits an important contrast in comparison
with the optical morphologies of other MMSNRs. In general, the optical morphologies of
MMSNRs resemble their radio morphologies in that they are shell-like in appearance:
examples of well-known MMSNRs with shell-like optical morphologies include CTB 1 
(G116.9$+$0.2) and CTA 1 (G119.5$+$10.2) (see \citet{Fesen97} and 
\citet{Mavromatakis00}, respectively). As described previously, the optical morphology of W28
is not shell-like but instead center-filled, exhibiting central diffuse ``puffy" structure: in
this sense, the optical morphology resembles that of another MMSNR -- HB21 (G89.0$+$4.7) 
-- that features both shell-like filamentary structure that is coincident with radio emission along
with patchy emission toward the SNR interior \citep{Mavromatakis07}. Like HB21, the
optical morphology of W28 may be sculpted by a complex interstellar medium into which
the SNR is expanding. 

\section{Discrete X-ray Sources}
\label{DiscreteXraySourceSection}

We now present a discussion about the two most prominent X-ray sources in the W28 field and
investigate whether either could be a neutron star associated with this SNR.  Searches for 
compact objects in MMSNRs are of
particular interest because so far only two of the 25 MMSNRs listed in Table 4 of \cite{Vink12} have firmly
associated neutron stars.\footnote{The specific MMSNRs are W44 and IC 443 with their associated neutron stars 
PSR B1853+01 and CXOU J061705.3$+$222127, respectively \citep{Petre02,Swartz15}.}  Hence, either most of 
these SN explosions leave no compact object, which seems
unlikely if these SNRs had massive stellar progenitors as commonly assumed, or the compact objects are
isolated, non-accreting black holes. 
The latter scenario seems more likely for SNRs in molecular clouds, since their massive progenitors could have 
had lifetimes shorter than those of the clouds that produced them.  Establishing the 
number of SNRs with black hole compact objects provides constraints on stellar
evolution theory, supernova explosion mechanisms and the mass function for massive stars. To date,
no neutron star has been conclusively associated with W28: the pulsar PSR B1758$-$23 \citep{Manchester85}
was once thought to be associated with the SNR \citep{Frail93} but this associated was later discarded based on VLBA 
measurements of the interstellar scattering in the direction of the pulsar \citep{Claussen99} along with a
large observed dispersion of the pulsar signal that cannot be reconciled easily with independent distance
estimates to W28 \citep{Kaspi93}.
\par
The first source of particular interest to us -- CXOU J175857.55$-$233400.3 --
appears to be the source of the hard X-ray emission detected from the southwest portion of
W28 by the prior {\it ASCA} observations of the SNR that were presented by \citet{Rho02}.
Those authors noted that the hard X-ray emission detected from W28 seemed to be localized
to this portion of the SNR; the {\it Chandra} observation presented here seems to favor a resolved source
strongly smeared out by the broad off-axis PSF. The other X-ray source --
3XMM J180058.5$-$232735 -- is the brightest X-ray source near the center of W28. The locations
of both of these sources are indicated in Figure \ref{W28Chandra3Color}.

\subsection{CXOU J175857.55$-$233400.3}
\label{CXOUJSubSection}

This source was imaged by the {\sl Chandra} ACIS  nearly $22'$ off-axis, and 
hence an analysis of the properties of this source is severely impacted by the PSF degradation. 
CXOU J175857.55$-$233400.3 may be  a true point source 
smeared out due to the broadened PSF and embedded in faint diffuse emission from the SNR,  which makes it appear more asymmetric 
and different in shape from what might be expected 
from simple PSF broadening. 
Alternatively, the extended emission may be intrinsic to the source. 
The peak in brightness of the source is located at 
R.A.=$17^{\rm h}58^{\rm m}57^{\rm s}.55$ and Decl.=$-23^{\circ}34'00~3''$ with a $1\sigma$ uncertainty of  $\approx15''$ (estimated 
as in \citet{Pavlov09}).   We extracted 1770 photons from a circle with a radius $r=22''$ 
(corresponding to the encircled source fraction of about 75\%) centered at the peak of the source surface brightness: of these counts, 54\% come from the surrounding background. 
The spectrum is shown in Figure \ref{PointSourceSpectra} (left panel): a statistically-acceptable
fit ($\chi_\nu$=1.05 for 14 degrees of freedom) is obtained using
an absorbed power law (PL) model with a photon index $\Gamma=0.47\pm0.25$ and a hydrogen
absorption column density\footnote{We used the {\tt phabs} model from XSPEC.} 
$N_H=(0.9\pm0.4$)$\times$10$^{22}$ cm$^{-2}$. The absorbed and unabsorbed fluxes for the
source are (4.1$\pm$0.2)$\times$10$^{-13}$ ergs cm$^{-2}$ s$^{-1}$ and 
$\approx4.7\times10^{-13}$ erg cm$^{-2}$ s$^{-1}$, respectively. 

We cross-correlated the position of 
CXOU J175857.55$-$233400.3 with those of  IR sources within a $15''$ distance using standard Vizier catalogs.  There are a total of 56 sources within this radius in the UKIDSS-DR6 catalog (with J-magnitude range of $\approx13-20$; a J=13 mag source is $11''$ away from the X-ray
source position) and 23 {\sl Spitzer} IRAC sources (IRAC band 1 magnitudes range is $\approx 11-14.5$; the nearest source is $4.5''$ away from the X-ray source position).  The closest counterpart (UGPS J175857.43$-$233357.4) found was in the UKIDSS-DR6 catalog at a distance of  $1.8''$ from CXOU J175857.55$-$233400.3, while the next closest UKIDSS source 
is $2.6''$ away and is dimmer by about 1 mag; others are beyond $4''$ away.  Trying to classify CXOU J175857.55--233400.3 by assuming all possible combinations of  IR and NIR counterparts within $15''$ radius would be a very laborious 
task. To obtain much more certain results, it is necessary to conduct a short  ($\sim10$ ks) on-axis {\sl Chandra} observation to pin down the position of the X-ray source to arcsecond precision. 
\par
Nonetheless, if we assume that the nearest UKIDSS  source (with magnitudes  $J=18.5\pm0.1$, $17.2\pm0.1$, and $16.7\pm 0.1$; not detected in IR) is the actual counterpart to CXOU 
J175857.5$-$233400.3, then this source would have a very large $F_X/F_{\rm NIR}$ ratio for a coronally active star (see Figure \ref{ColorColorFigures};  these conclusions would not change if the second nearest UKIDSS source were the actual counterpart). A Wolf-Rayet (non-binary) star 
would have a much softer X-ray spectrum than that extracted for this source and it would
appear to be bright in the infrared (specifically in the {\sl Spitzer} IRAC bands) at the extinction inferred from X-ray spectral fits that place CXOU J175857.5$-$233400.3  closer than the Galactic center. The 
nondetection of UGPS J175857.43$-$233357.4 
by the {\sl Spitzer} IRAC (at the 8 micron band) and MIPS (at the 24 micron band) suggests that the source is not a 
young stellar object (YSO) or protostar. Such objects usually cluster together, and we do not see other X-ray sources with similar X-ray spectra within the immediate vicinity.  Confident identification of this X-ray source with    a NIR counterpart 
would further exclude  classification
of the source as an isolated neutron star -- which is already unlikely given the very hard spectrum of the source.
The remaining possible classifications of this X-ray source are as a cataclysmic variable (CV) or
a quiescent low-mass X-ray binary.  These are supported by the location of CXOU J175857.55--233400.3 in the diagrams shown in Figure \ref{ColorColorFigures}.  Emission is detected at a longer wavelength (0.87 mm)
in an ATLASGAL (0.87 mm) image at a location north of CXOU J175857.55$-$233400.3  and at 
radio wavelengths (20 cm and 90 cm) larger-scale regions of enhanced emission stretching from north to south are detected. In Figure \ref{HardSourceMultiWavelengthFigures} we present
multiwavelength (chiefly infrared and radio) images of the neighborhood surrounding CXOU J175857.55$-$233400.3.

\subsection{3XMM J180058.5$-$232735}
\label{3XMMSubSection}

This X-ray source is of a particular interest because it is located close to the geometric center of W28
and is relatively bright. Although it lies outside the {\em Chandra} footprint, it is seen in several {\it XMM-Newton} observations listed in the 3XMM-DR5 catalog \citep{Rosen16}, but always 
relatively far off-axis ($12'$--$15'$). Nevertheless, it has a well-determined position with $1\sigma$ uncertainty of $0.6''$ in 
the 3XMM-DR5 catalog. In this paper, we will adopt the coordinates for this source to be 
R.A.=$18^{\rm h}00^{\rm m}58^{\rm s}.59$ and Decl.=$-23^{\circ}27'35~9''$ (J2000.).  There is a fairly bright optical/NIR/IR counterpart detected in the  only $\approx0.4''$ from the 3XMM position. The UCAC4 Catalog \citep{Zacharias13} lists magnitudes for this star (UCAC4-333-134563) as B=14.45, V=13.44, R=13.11 and I=12.69, while UKIDSS-DR6 magnitudes are 
$J= 11.23$, $H=11.02$, and $K=10.66$. Finally, the infrared counterpart detected by the WISE all-sky survey has
magnitudes $W1=9.80$, $W2=9.80$, and  $W3= 9.27$. The colors indicated by these counterparts are used to plot the location of 3XMM J180058.5$-$232735 in the diagrams shown in Fig.~\ref{ColorColorFigures}.
The presence of a bright counterpart to the X-ray source at optical, near-infrared and infrared
wavelengths -- coupled with the locations of the X-ray source within the color-color diagrams --
tend to rule out  classification of 3XMM J180058.5$-$232735 as a background AGN or
a neutron star or pulsar. The most likely classification of this X-ray source is as a relatively nearby
foreground star with an active corona. This classification is supported by the large proper motion of UCAC4-333-134563: 
$\Delta$RA$=-24.2\pm2.0$ mas yr$^{-1}$ and $\Delta$Decl$=11.5\pm5.6$ mas yr$^{-1}$.  
Furthermore, the automated classification algorithm for X-ray sources described by 
\citet{Sonbas16} and \citet{Hare16} favors a classification of the source as either a coronally-active main-sequence (MS) star or a YSO.  
\par
We fitted an absorbed power law model to the source spectrum (extracted from XMM EPIC-PN data for ObsID 0135742201). The spectrum is shown in Figure \ref{PointSourceSpectra} (right panel). The best-fit absorbed PL  parameters are $N_H=(3.2\pm1.5)\times10^{21}$~cm$^{-2}$ and 
$\Gamma=3.8\pm1.3$. The spectrum is soft and consistent with active coronae spectra: the small number of
counts and low signal-to-noise preclude any more detailed spectral fitting. 
	
\section{Summary and Conclusions}
\label{ConclusionsSection}

The conclusions of this paper may be described as follows:
\par
1) We present and analyze for the first time optical (CTIO) and X-ray ({\it Chandra} ACIS and  
{\it ROSAT} HRI) imaging observations of the archetypical mixed-morphology
SNR W28. 
These observations have allowed us to analyze X-ray emission
from this SNR with the highest angular resolution yet (approximately
$12''$ for our {\it ROSAT} HRI mosaicked image of W28 and as high as
$1''$ for the portions of W28 sampled by {\it Chandra}.)
\par
2) The new narrowband [\ion{S}{2}] and H$\alpha$ optical images of W28 reveal
extensive diffuse structure with patchy knots and filaments. The [\ion{S}{2}] emission is strongest
relative to the H$\alpha$ emission along the northern and eastern rims of the SNR;
large numbers of OH masers are seen along these same rims and thus indicate
that shock interactions taking place between W28 and adjacent molecular clouds at
these sites. The new {\it ROSAT} HRI mosaic image of W28 shows very starkly the
centrally-concentrated X-ray morphology of the SNR and its strong anti-correlation
with the shell-like radio morphology. 
\par
3) We have performed a spatially-resolved spectroscopic analysis of the {\it Chandra} 
observation of W28 where we have analyzed the spectra extracted from the bright central
peak, several arcminute-scale regions near the central peak and annular regions centered
on the bright peak. We have fit the spectrum of the peak with two VRNEI components:
the fitted elemental abundances are subsolar, indicating that the emission from the peak
is dominated by swept-up material rather than ejecta. Fits to the arcminute-scale regions
with two VRNEI components produce generally consistent results with the fit to the spectra
of the central peak region: in these fits, we find evidence for enhanced oxygen abundance
relative to iron abundances. This result is expected
for SNRs with massive stellar progenitors and -- together with the evidence for an association with 
adjacent molecular clouds -- further cements the interpretation that
W28 arose from a massive stellar progenitor. Finally, we have fit spectra from concentric
annular regions centered on the bright central peak to search for gradients in temperature
and elemental abundances: consistent with other MMSNRs, we find no evidence for
a temperature gradient with increasing distance from the bright central peak. In contrast
to prediction, we find no evidence for a gradient in elemental abundance either. 
\par
4) We have presented a comparison of the spatial morphologies of W28 at X-ray, optical and
radio wavelengths. As the archetypical MMSNR, W28 features a shell-like radio morphology
with a contrasting center-filled X-ray morphology with a thermal spectrum.   The only spatial feature that is detected at all three wavelengths is the northeastern
``ear"; this region appears to be a site of a particularly dramatic interaction between the W28 shock and
an adjacent molecular cloud. In contrast to other MMSNRs, the optical morphology of W28
is not shell-like but instead center-filled; this may be due to the particularly complex 
interstellar medium conditions into which W28 is expanding. 
\par
5) We have analyzed the X-ray properties of two individual X-ray sources that stand out in the X-ray images:
CXOU J175857.55$-$233400.3 (a hard X-ray source seen toward the 
southwest portion of W28 by 
prior {\it ASCA} observations of the SNR) and 3XMM
J180058.5$-$232735 (a point-like X-ray source seen near the geometric
center of the SNR). These sources have attracted interest as candidate X-ray-emitting
neutron stars that may be associated with W28. Through an analysis of spectral
properties (including fits to extracted X-ray spectra) and a search for counterparts
at different wavelengths, we conclude that neither of these sources is a plausible candidate 
for a neutron star associated with W28. The former source is most likely to be a CV or a 
quiescent LMXB while the latter is most likely to be a coronally-active MS star.  

\acknowledgments

We thank the anonymous referee for many useful comments that improved the quality of this manuscript.
We thank G. Dubner for kindly sharing her radio image of W28 and Kazik Borkowski for helpful
discussions about X-ray emission from SNRs. T.G.P. thanks Joseph DePasquale and the Chandra
X-ray Center Help Desk Team in preparing the exposure-corrected {\it Chandra} images of W28.
T.G.P. also thanks Steven Snowden
for his assistance in preparing the HRI mosaicked image of W28. 
This research has made use of data obtained from the High 
Energy Astrophysics Science Archive Research Center (HEASARC), provided by
NASA's Goddard Space Flight Center. This research has also made use of NASA's 
Astrophysics Data System Bibliographic Services. PFW acknowledges support from the NSF under grant AST-0908566.

\clearpage

\begin{deluxetable}{lccc}
\tablecaption{Summary of Optical Imaging Observations of W28\label{OpticalObsTable}}
\tablewidth{0pt}
\tablehead{
\colhead{} & \multicolumn{2}{c}{Filter} &   \colhead{Exposure}   \\
\colhead{Designation} & 
\colhead{$\lambda_0$\,(\AA)} &\colhead{$\Delta\lambda$\tablenotemark{a}(\AA)}   &\colhead{ (s)} 
}
\startdata
\ha &6568  &  28 & $4  \times 600 $ \\
\sii\ \LL 6716, 6731 &6728  & 50  & $4  \times 600 $ \\
Red Continuum &6852  &  95  & $4  \times 300 $ \\
\enddata
\tablenotetext{a}{Full width at half maximum.}
\end{deluxetable}

\clearpage


\begin{deluxetable}{cccccccc}
\tablecaption{Summary of {\it ROSAT} HRI and {\it Chandra} ACIS 
Observations of W28\label{Table1}}
\tabletypesize{\scriptsize}
\tablewidth{0pt}
\tablehead{
& & & \colhead{Nominal} & \colhead{Nominal} & & \colhead{Effective}\\
& & & \colhead{R.A.} & \colhead{Decl.} & & \colhead{Exposure} 
& \colhead{Sampled}\\
\colhead{Observatory and} & \colhead{Sequence} & & \colhead{(J2000.0)} & 
\colhead{(J2000.0)} & & \colhead{Time\tablenotemark{b}} & \colhead{Portion}\\
\colhead{Instrument} & \colhead{Number} & \colhead{ObsID\tablenotemark{a}} & 
\colhead{(h m s)} & \colhead{($\circ$ ' ")} & \colhead{Date} & 
\colhead{(seconds)} & \colhead{of W28}
}
\startdata
{\it ROSAT} HRI & RH500382N00 & -- & 18 00 48.0 & $-$23 20 24.0 &  
20 March 1995 & 23765 & Center \\
{\it ROSAT} HRI & RH500382A01 & -- & 18 00 48.0 & $-$23 20 24.0 & 
27 March 1996 & 11319 & Center \\
{\it ROSAT} HRI & RH500484N00 & -- & 18 01 19.2 & $-$23 04 48.0 & 
11 September 1998 & 27120 & Northeast \\
{\it ROSAT} HRI & RH500485N00 & -- & 17 59 14.4 & $-$23 19 12.0 & 
5 October 1997 & 15134 & Northwest \\
{\it ROSAT} HRI & RH500486N00 & -- & 18 01 43.2 & $-$23 35 24.0 & 
15 March 1998 & 3992 & Southeast \\
{\it ROSAT} HRI & RH500486A01 & -- & 18 01 43.2 & $-$23 35 24.0 &
11 September 1998 & 14856 & Southeast \\
{\it ROSAT} HRI & RH500487N00 & -- & 18 00 19.2 & $-$23 36 00.0 & 
14 March 1998 & 14211 & South \\
{\it ROSAT} HRI & RH500487A01 & -- & 18 00 19.2 & $-$23 36 00.0 &
8 September 1998 & 15300 & South \\
{\it ROSAT} HRI & RH500488N00 & -- & 17 59 00.0 & $-$23 36 00.0 & 
8 September 1998 & 6576 & Southwest \\
{\it Chandra} ACIS & 500278 & 2828 & 18 00 24.6 & $-$23 25 55.7 & 
12 October 2002 & 88994 & Center and\\
& & & & & & & Southwest
\enddata
\tablenotetext{a}{For the {\it Chandra} ACIS observation only.}
\tablenotetext{b}{In the cases of the {\it ROSAT}~HRI observations, the 
given exposure times are deadtime-corrected. Broken exposures (RH500382N00 
and RH500382A01, RH500486N00 and RH500486A01, RH500487N00 and RH500487A01) 
have been combined into single datasets for the purposes of boosting the 
signal-to-noise and preparing a mosaicked {\it ROSAT} HRI image of 
W28. See Section \ref{HRIObsSection}.}
\end{deluxetable}

\clearpage

\begin{deluxetable}{lcccc}
\tablecaption{Summary of Fits to Extracted {\it Chandra} Spectrum of Central 
Region\tablenotemark{a}\label{BlobSpectrumTable}}
\tabletypesize{\scriptsize}
\tablewidth{0pt}
\tablehead{
& \colhead{TBABS$\times$} & \colhead{TBABS$\times$} & \colhead{TBABS$\times$} &
\colhead{TBABS$\times$}\\
\colhead{Parameter} & \colhead{(VAPEC+VAPEC)} & \colhead{(VNEI+VNEI)} & \colhead{VRNEI}
& \colhead{(VRNEI+VRNEI)}
}
\startdata
$N$$_H$ (10$^{22}$ cm$^{-2}$) & 0.43$\pm$0.02 & 0.41$^{+0.03}_{-0.02}$ & 0.50$\pm$0.02
& 1.14$^{+0.09}_{-0.12}$ \\
$kT$$_{\rm{e1}}$ (keV) & 0.65$\pm$0.01 & 0.68$\pm$0.01 & 0.61$^{+0.02}_{-0.01}$ & 
0.10$\pm$0.01\\
$kT$$_{\rm{Z1}}$ (keV) & \nodata & \nodata & 10.31$^{+13.94}_{-4.02}$ & 0.16$^{+0.02}_{-0.01}$ \\
$\tau$$_{\rm{1}}$ (10$^{11}$ cm$^{-3}$ s) & \nodata & 4.51$^{+0.42}_{-0.51}$ &
6.24$^{+0.42}_{-0.51}$ & $<$1.24 \\
Normalization$_{\rm{1}}$\tablenotemark{b} (cm$^{-5}$) & 3.38$\times$10$^{-3}$ 
& 2.69$\times$10$^{-3}$ & 7.41$\times$10$^{-3}$ & 0.60 \\
$kT$$_{\rm{e2}}$ (keV) & 1.70$^{+0.17}_{-0.14}$ & 1.81$^{+0.39}_{-0.18}$ & \nodata & 
0.42$^{+0.05}_{-0.04}$ \\
$kT$$_{\rm{Z2}}$ (keV) & \nodata & \nodata & \nodata & 3.08$^{+2.86}_{-0.97}$ \\
$\tau$$_{\rm{2}}$ (10$^{11}$ cm$^{-3}$ s) & \nodata & 5.75$^{+7.12}_{-3.06}$ & \nodata & 
5.55$^{+0.49}_{-0.63}$\\
Normalization$_{\rm{2}}$\tablenotemark{b} (cm$^{-5}$) & 7.79$\times$10$^{-4}$ & 
8.31$\times$10$^{-4}$ & \nodata & 1.65$\times$10$^{-2}$ \\
\hline
Abundances\tablenotemark{c} \\
\hline 
O & 1 (frozen) & 1.32$^{+0.25}_{-0.26}$ & 0.55$^{+0.15}_{-0.14}$ & 0.62$^{+0.26}_{-0.19}$ \\
Ne & 1.44$^{+0.11}_{-0.14}$ & 1 (frozen) & 0.54$\pm$0.10 & 0.59$^{+0.14}_{-0.13}$ \\
Mg & 1 (frozen) & 1 (frozen) & 0.72$^{+0.05}_{-0.07}$ & 1 (frozen) \\
Fe & 0.45$\pm$0.02 & 0.51$^{+0.04}_{-0.03}$ & 0.33$^{+0.04}_{-0.03}$ & 
0.40$^{+0.07}_{-0.05}$ \\
\hline
$\chi$$^2$/DOF & 326.18/180 & 316.07/178 & 310.10/178 & 205.00/175\\
$\Delta$$\chi$$^2$ & 1.81 & 1.78 & 1.74 & 1.17 \\
\enddata
\tablenotetext{a}{For the energy range 0.5-4.0 keV. All quoted errors are at the 90\% confidence
level.}
\tablenotetext{b}{Defined in units of (10$^{-14}$/4$\pi$$d$$^2$)$\int$$n$$_e$$n$$_p$$d$$V$,
where $d$ is the distance to the source (in cm), $n$$_e$ and $n$$_p$ are the number densities
of electrons and hydrogen nuclei, respectively (in units of cm$^{-3}$) and finally $\int$$d$$V$ is
the integral over the entire volume of the X-ray-emitting plasma (in units of cm$^{-3}$).}
\tablenotetext{c}{Relative to solar.}

\end{deluxetable}

\clearpage

\begin{deluxetable}{lcccccc}
\tablecaption{Summary of Fits to Extracted Spectra of Five ACIS-S3 Regions and ACIS-S4 
Region with TBABS$\times$VNEI Model\tablenotemark{a}\label{GroupVNEITable}}
\tabletypesize{\scriptsize}
\tablewidth{0pt}
\tablehead{
& & & & & &  \colhead{ACIS-S4} \\
\colhead{Parameter} & \colhead{Region 1} & \colhead{Region 2} &
\colhead{Region 3} & \colhead{Region 4} & \colhead{Region 5} & \colhead{Region}}
\startdata
$k$$T$$_{\rm{e}}$ (keV) & 0.64$^{+0.01}_{-0.02}$ & 0.75$^{+0.03}_{-0.04}$
& 0.69$^{+0.01}_{-0.02}$ & 0.67$^{+0.03}_{-0.02}$ & 0.69$\pm$0.01 & 0.66$\pm$0.01\\
O\tablenotemark{b} & 3.55$^{+0.95}_{-0.79}$ & 4.59$^{+1.22}_{-1.63}$ & 4.67$^{+1.42}_{-1.28}$
& 6.79$^{+1.12}_{-1.03}$ & 8.69$^{+0.68}_{-0.67}$ & 4.28$^{+0.77}_{-0.82}$ \\
Fe\tablenotemark{b} & 0.59$\pm$0.06 & 0.78$^{+0.17}_{-0.07}$ & 0.71$^{+0.10}_{-0.14}$  
& 1.00 (frozen) & 1.00 (frozen) & 0.57$^{+0.05}_{-0.06}$\\
$\tau$ (10$^{11}$ cm$^{-3}$ s$^{-1}$) & $>$10 & 4.24$^{+1.22}_{-1.09}$ & 4.68$^{+1.11}_{-0.75}$ 
& 4.43$^{+1.24}_{-0.96}$ & 3.82$^{+0.43}_{-0.49}$ & 4.31$^{+0.70}_{-0.27}$ \\
Normalization\tablenotemark{c} (cm$^{-5}$) & 8.09$\times$10$^{-4}$ & 3.14$\times$10$^{-4}$  
& 5.91$\times$10$^{-4}$ & 3.11$\times$10$^{-4}$ & 1.14$\times$10$^{-3}$ & 
2.89$\times$10$^{-3}$ \\
\enddata
\tablenotetext{a}{All spectra were extracted over the energy range 0.5-4.0 keV except for the
ACIS-S4 region, which was extracted over the energy range 0.6-2.0 keV.  
All quoted errors are at the 90\% confidence level. The column densities for each spectrum were
tied together with a fitted value of $N$$_{\rm{H}}$ = 0.65$\pm$0.02$\times$10$^{22}$ 
cm$^{-2}$. For this fit, $\chi$$^2$/DOF = $\Delta$$\chi$$^2$ = 1077.45/622 = 1.73.}
\tablenotetext{b}{Relative to solar.}
\tablenotetext{c}{Defined in units of (10$^{-14}$/4$\pi$$d$$^2$)$\int$$n$$_e$$n$$_p$$d$$V$,
where $d$ is the distance to the source (in cm), $n$$_e$ and $n$$_p$ are the number densities
of electrons and hydrogen nuclei, respectively (in units of cm$^{-3}$) and finally $\int$$d$$V$ is
the integral over the entire volume of the X-ray-emitting plasma (in units of cm$^{-3}$).}
\end{deluxetable}

\clearpage


\begin{deluxetable}{lcccccc}
\tablecaption{Summary of Fits to Extracted Spectra of Five ACIS-S3 Regions and ACIS-S4
Region with TBABS$\times$VRNEI Model\tablenotemark{a}\label{GroupFitVRNEITable}}
\tabletypesize{\scriptsize}
\tablewidth{0pt}
\tablehead{
& & & & & & \colhead{ACIS-S4} \\
\colhead{Parameter} & \colhead{Region 1} & \colhead{Region 2} &
\colhead{Region 3} & \colhead{Region 4} & \colhead{Region 5} & \colhead{Region}}
\startdata
$k$$T$$_{\rm{e}}$\tablenotemark{b} (keV) & 0.65$\pm$0.01 & 0.67$\pm$0.04 & 
0.63$\pm$0.02 & 0.68$\pm$0.03 & 0.63$\pm$0.02 & 0.62$\pm$0.01\\
$k$$T$$_{\rm{Z}}$\tablenotemark{c} (keV) & $>$5 & 7.14 ($>$2.54) & 6.46 ($>$2.17) & 
$>$5 & 5.11$^{+8.78}_{-2.71}$ & $>$5 \\
O\tablenotemark{d} & 2.47$^{+0.73}_{-0.57}$ & 1.00 (frozen) & 1.00 (frozen) &  
3.28$^{+1.69}_{-1.05}$ & 2.05$^{+0.45}_{-0.37}$ & 1.48$^{+0.27}_{-0.30}$ \\
Fe\tablenotemark{d} & 0.46$^{+0.05}_{-0.04}$ & 0.45$^{+0.10}_{-0.08}$ & 
0.39$^{+0.07}_{-0.04}$ & 0.65$^{+0.13}_{-0.09}$ & 0.49$^{+0.08}_{-0.04}$ & 
0.36$^{+0.02}_{-0.04}$ \\
$\tau$ (10$^{11}$ cm$^{-3}$ s$^{-1}$) & $>$10$^{12}$ & 6.95$^{+2.22}_{-3.87}$ & 
7.86$^{+2.10}_{-1.63}$ & $>$10$^{12}$ & 6.94$^{+1.19}_{-1.08}$ & $>$10$^{12}$ \\ 
Normalization\tablenotemark{e} (cm$^{-5}$) & 8.25$\times$10$^{-3}$ & 5.36$\times$10$^{-4}$
& 9.99$\times$10$^{-4}$ & 4.03$\times$10$^{-4}$ & 2.24$\times$10$^{-3}$ & 
4.01$\times$10$^{-3}$ \\ 
\enddata
\tablenotetext{a}{The extracted spectra for the annular regions located on the ACIS-S3 chip 
were fitted over the energy range 0.5-4.0 keV while the extracted spectrum for the region 
on the ACIS-S4 chip were fitted over
the energy range 0.6-2.0 keV. All quoted errors are at the 90\% confidence
level. The column densities for each spectrum were
tied together with a fitted value of $N$$_{\rm{H}}$ = 0.57$\pm$0.01$\times$10$^{22}$ 
cm$^{-2}$. For this fit, $\chi$$^2$/DOF = $\Delta$$\chi$$^2$ = 1024.64/616 = 1.66.}
\tablenotetext{b}{The electron temperature.}
\tablenotetext{c}{The ion temperature.}
\tablenotetext{d}{Relative to solar.}
\tablenotetext{e}{Defined in units of (10$^{-14}$/4$\pi$$d$$^2$)$\int$$n$$_e$$n$$_p$$d$$V$,
where $d$ is the distance to the source (in cm), $n$$_e$ and $n$$_p$ are the number densities
of electrons and hydrogen nuclei, respectively (in units of cm$^{-3}$) and finally $\int$$d$$V$ is
the integral over the entire volume of the X-ray-emitting plasma (in units of cm$^{-3}$).}
\end{deluxetable}

\clearpage

\begin{deluxetable}{lcccccc}
\tablecaption{Summary of Fits to Extracted Spectra of Five ACIS-S3 Regions and ACIS-S4
Region with TBABS$\times$(VRNEI+VRNEI) Model \label{SixRegionsVRNEIVRNEIFitsTable}}
\tabletypesize{\scriptsize}
\tablewidth{0pt}
\tablehead{
& & & & & & \colhead{ACIS-S4} \\
\colhead{Parameter} & \colhead{Region 1} & \colhead{Region 2} &
\colhead{Region 3} & \colhead{Region 4} & \colhead{Region 5} & \colhead{Region}
}
\startdata
$kT$$_{\rm{e1}}$\tablenotemark{b} (keV) & 0.19$\pm$0.02 & 0.08 ($>$0.10) & 
0.19$^{+0.08}_{-0.04}$ & 0.08 ($<$0.10) & 0.21$^{+0.01}_{-0.02}$ & 0.41$^{+0.07}_{-0.08}$ \\
$kT$$_{\rm{Z1}}$\tablenotemark{c} (keV) & 1.83($>$1.12) & 0.19$^{+0.19}_{-0.02}$ &
$>$5 & $>$5 &$>$2.90 & 0.74$\pm$0.06 \\
$\tau$$_{\rm{1}}$ (10$^{11}$ cm$^{-3}$ s) & 7.49($\pm$0.94) & $<$10 & $>$10
& 10.29$^{+1.75}_{-1.70}$ & 6.76$^{+0.60}_{-0.67}$ & $<$0.01  \\
Normalization$_{\rm{1}}$\tablenotemark{d} (cm$^{-5}$) & 6.27$\times$10$^{-3}$ & 
6.46$\times$10$^{-3}$ & 3.68$\times$10$^{-4}$ & 1.22$\times$10$^{-2}$ &
7.69$\times$10$^{-3}$ & 1$\times$10$^{-3}$\\
$kT$$_{\rm{e2}}$\tablenotemark{b} (keV) & 0.60$^{+0.08}_{-0.10}$ & 0.56$^{+0.06}_{-0.08}$ 
& 0.64$^{+0.01}_{-0.02}$ & 0.35$^{+0.05}_{-0.04}$ & 0.67$^{+0.05}_{-0.04}$ &
0.34$^{+0.02}_{-0.01}$ \\
$kT$$_{\rm{Z2}}$\tablenotemark{c} (keV) & 3.76$^{+7.81}_{-1.89}$ & 3.66 ($>$1.39) & $>$5 &
2.08$^{+36.87}_{-0.55}$ & 3.49$^{+3.50}_{-1.46}$ & $>$5\\
$\tau$$_{\rm{2}}$ (10$^{11}$ cm$^{-3}$ s) & 5.19$^{+2.27}_{-1.30}$ & 5.74$^{+1.79}_{-2.26}$ 
& $>$10 & 2.77$^{+1.99}_{-0.75}$ & 5.81$^{+1.42}_{-1.08}$ & $>$10 \\
Normalization$_{\rm{2}}$\tablenotemark{d} (cm$^{-5}$) & 7.90$\times$10$^{-4}$ & 
8.03$\times$10$^{-4}$ & 7.44$\times$10$^{-4}$ &1.16$\times$10$^{-3}$ & 1.29$\times$10$^{-3}$ 
& 6.21$\times$10$^{-3}$ \\
\hline
Abundances\tablenotemark{e} \\
\hline 
O & 0.48$^{+0.31}_{-0.18}$ & 1.00 (frozen) & 3.60$^{+0.89}_{-0.80}$ & 1.00 (frozen) &
1.00 (frozen) & 1.38$^{+0.35}_{-0.27}$ \\
Ne & 0.49$^{+0.29}_{-0.24}$ & 1.00 (frozen) & 2.08$^{+0.40}_{-0.37}$ & 0.67$^{+0.31}_{-0.22}$ 
& 1.00 (frozen) & 0.64$^{+0.16}_{-0.17}$ \\
Fe & 1.00 (frozen) & 0.40$^{+0.16}_{-0.10}$ & 1.00 (frozen) & 1.00 (frozen) &
1.23$^{+0.31}_{-0.20}$ & 0.47$\pm$0.06 \\
\enddata
\tablenotetext{a}{The extracted spectra for the regions located on the ACIS-S3 chip were fitted over 
the energy range 0.5-4.0 keV while the extracted spectrum for the region on the ACIS-S4 chip were 
fitted over
the energy range 0.6-2.0 keV. All quoted errors are at the 90\% confidence
level. The column densities for each spectrum were
tied together with a fitted value of $N$$_{\rm{H}}$ = 0.88$\pm$0.03$\times$10$^{22}$ 
cm$^{-2}$. For this fit, $\chi$$^2$/DOF = $\Delta$$\chi$$^2$ = 734.87/692 = 1.24.}
\tablenotetext{b}{The electron temperature.}
\tablenotetext{c}{The ion temperature.}
\tablenotetext{d}{Defined in units of (10$^{-14}$/4$\pi$$d$$^2$)$\int$$n$$_e$$n$$_p$$d$$V$,
where $d$ is the distance to the source (in cm), $n$$_e$ and $n$$_p$ are the number densities
of electrons and hydrogen nuclei, respectively (in units of cm$^{-3}$) and finally $\int$$d$$V$ is
the integral over the entire volume of the X-ray-emitting plasma (in units of cm$^{-3}$).}
\tablenotetext{e}{Relative to solar.}
\end{deluxetable}

\clearpage

\begin{deluxetable}{lcccccccccc}
\rotate
\tablecaption{Summary of TBABS$\times$VRNEI Fits to Extracted Spectra for Annular Regions of 
ACIS-S3 Chip and ACIS-S4 Chip Region\tablenotemark{a}\label{AnnularRegionsTable}}
\tabletypesize{\scriptsize}
\tablewidth{0pt}
\tablehead{
& \colhead{Annulus} & \colhead{Annulus} & \colhead{Annulus} & \colhead{Annulus}
& \colhead{Annulus} & \colhead{Annulus} & \colhead{Annulus} & \colhead{Annulus}  & \colhead{S4 Chip}\\
\colhead{Parameter} & \colhead{1} & \colhead{2} & \colhead{3} & \colhead{4} & 
\colhead{5} & \colhead{6} & \colhead{7} & \colhead{8} & \colhead{Region}}
\startdata
$kT$$_{\rm{e}}$\tablenotemark{b} & 0.62$\pm$0.01 & 0.65$^{+0.02}_{-0.01}$ & 0.64$\pm$0.01
& 0.62$\pm$0.01 & 0.64$\pm$0.01 & 0.65$\pm$0.02 & 0.64$\pm$0.01 & 0.66$^{+0.02}_{-0.01}$ 
& 0.62$\pm$0.01\\

$kT$$_{\rm{Z}}$\tablenotemark{c} & 3.53$^{+2.68}_{-1.59}$ & 3.25$^{+7.92}_{-0.72}$ 
& $>$4.47
& 4.55$^{+2.53}_{-2.43}$ & 5.54$^{+2.57}_{-1.90}$ & 4.47$^{+2.21}_{-1.30}$ 
& 5.09$^{+2.28}_{-0.80}$ & 6.65$^{+4.24}_{-2.34}$ & $>$5.00\\

O\tablenotemark{d} & 1.00$^a$ & 1.81$^{+0.32}_{-0.37}$ & 
1.29$^{+0.11}_{-0.26}$ & 1.00 (frozen) & 1.20$^{+0.09}_{-0.20}$ & 1.52$^{+0.18}_{-0.19}$ & 
1.40$^{+0.12}_{-0.10}$ & 1.57$^{+0.32}_{-0.24}$ & 1.45$^{+0.22}_{-0.23}$ \\

Fe\tablenotemark{d} & 0.37$\pm$0.03 & 0.40$\pm$0.04 & 0.43$^{+0.05}_{-0.02}$ & 
0.35$\pm$0.02 & 0.36$^{+0.03}_{-0.02}$ & 0.37$^{+0.02}_{-0.03}$ & 0.34$^{+0.01}_{-0.02}$ & 
0.36$^{+0.03}_{-0.04}$ & 0.35$^{+0.01}_{-0.03}$\\

$\tau$ (10$^{11}$ cm$^{-3}$ s) & 6.21$^{+0.68}_{-1.07}$ & 6.40$^{+1.64}_{-0.78}$ 
& 7.45$^{+1.40}_{-0.97}$ & 6.14$^{+0.53}_{-0.44}$ & 7.15$^{+0.73}_{-0.70}$ 
& 6.95$^{+0.47}_{-0.54}$ & 6.98$^{+0.61}_{-0.30}$ & 7.44$^{+0.87}_{-0.66}$ & $>$20\\

Normalization\tablenotemark{e} (10$^{-3}$ cm$^{-5}$) & 1.65 & 2.13 & 2.86 & 3.90 & 4.14 & 
4.45 & 5.18 & 5.42 & 4.03

\enddata
\tablenotetext{a}{The extracted spectra for the annular regions located on the ACIS-S3 chip were fitted over 
the energy range 0.5-4.0 keV while the extracted spectrum for the region on the ACIS-S4 chip were fitted over
the energy range 0.6-2.0 keV. All quoted errors are at the 90\% confidence
level. The column densities for each spectrum were
tied together with a fitted value of $N$$_{\rm{H}}$ = 0.57$\pm$0.01$\times$10$^{22}$ 
cm$^{-2}$. For this fit, $\chi$$^2$/DOF = $\Delta$$\chi$$^2$ = 2394.02/1379 = 1.74.}
\tablenotetext{b}{The electron temperature.}
\tablenotetext{c}{The ion temperature.}
\tablenotetext{d}{Relative to solar.}
\tablenotetext{e}{Defined in units of (10$^{-14}$/4$\pi$$d$$^2$)$\int$$n$$_e$$n$$_p$$d$$V$,
where $d$ is the distance to the source (in cm), $n$$_e$ and $n$$_p$ are the number densities
of electrons and hydrogen nuclei, respectively (in units of cm$^{-3}$) and finally $\int$$d$$V$ is
the integral over the entire volume of the X-ray-emitting plasma (in units of cm$^{-3}$).}
\end{deluxetable}

\clearpage
\begin{figure}
\plotone{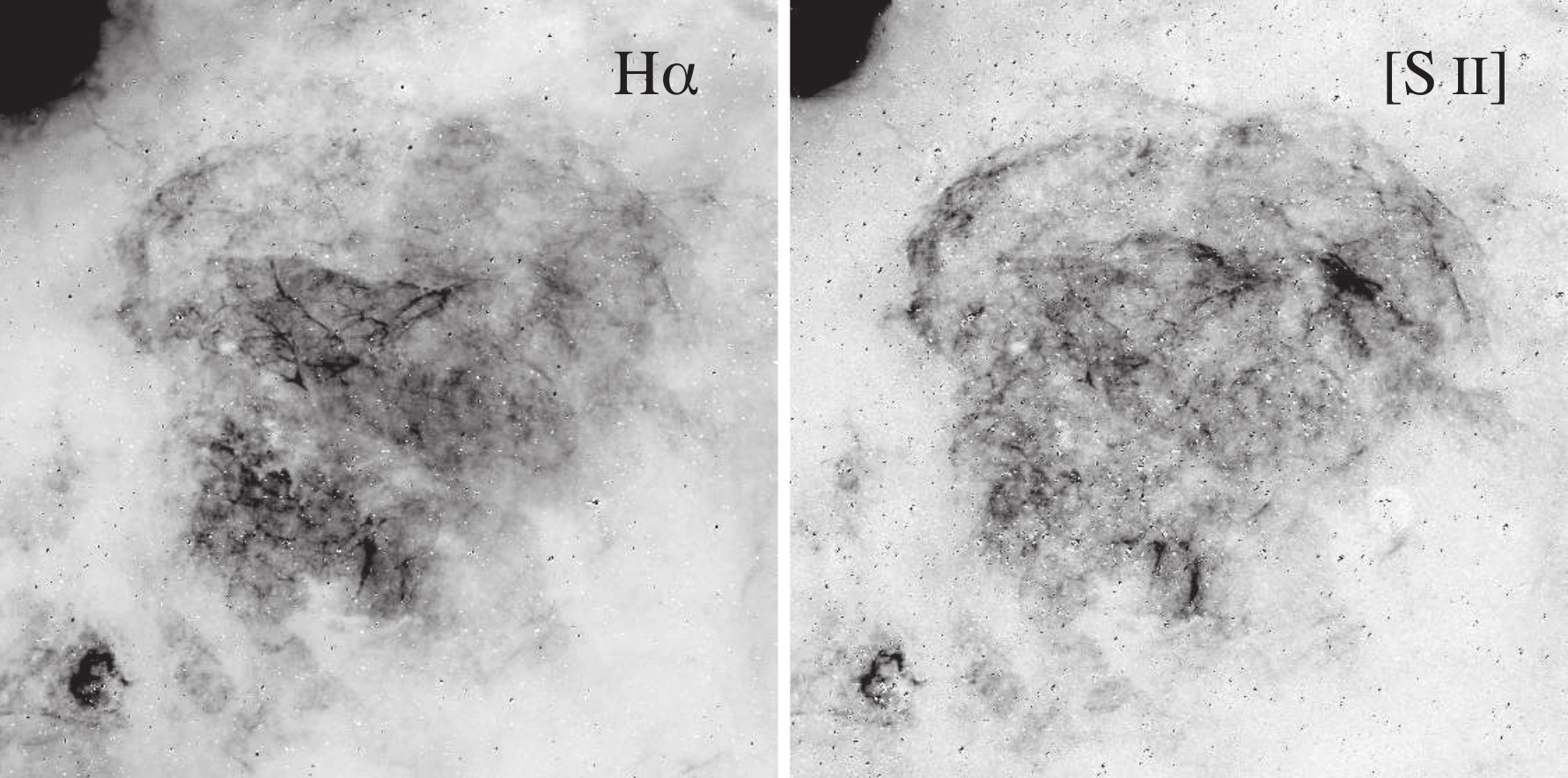}
\caption{Images of W28 in \ha\ (left) and \sii\ $\lambda\lambda$\,6716,6731 emission lines,
taken from the Burrell Schmidt 
telescope at KPNO.  A continuum image has been scaled and subtracted from each.  The field is 45\arcmin\ square, oriented north up and east left.  
The bright
source seen toward the upper left is a portion of M20 (the Trifid Nebula).}
\label{figure_ha_sii_pair}
\end{figure}

\clearpage
\begin{figure}
\includegraphics[scale=0.65,angle=-90]{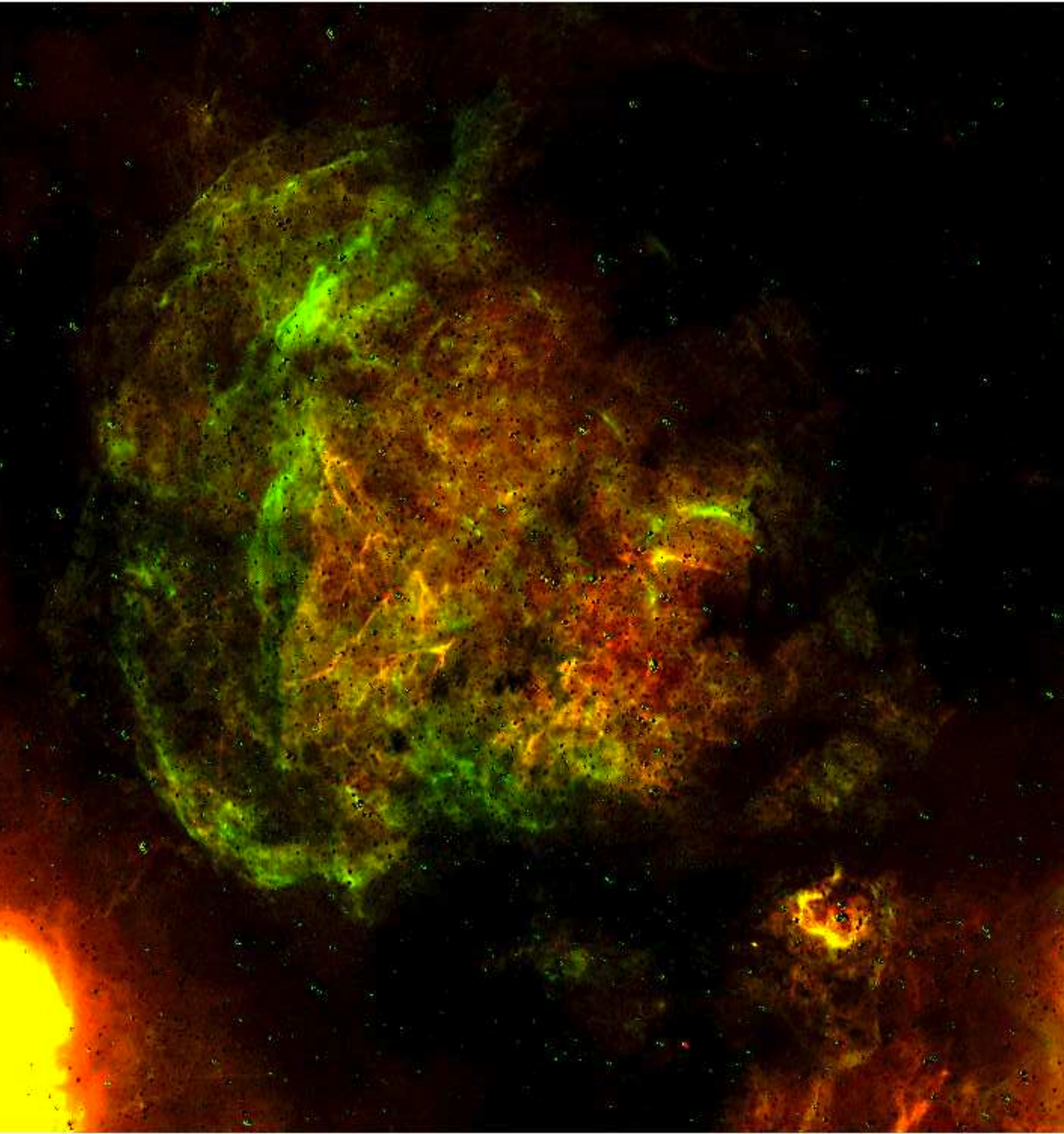}
\caption{Two-color optical image of W28: H$\alpha$ emission is shown in red, and \sii\ 
 in green. The continuum-subtracted images are the same as in Fig.~\ref{figure_ha_sii_pair}, but the field is somewhat larger, 50\arcmin\ square.}
\label{figure_w28_color}
\end{figure} 

\clearpage
\begin{figure}
\includegraphics[scale=0.65,angle=-90]{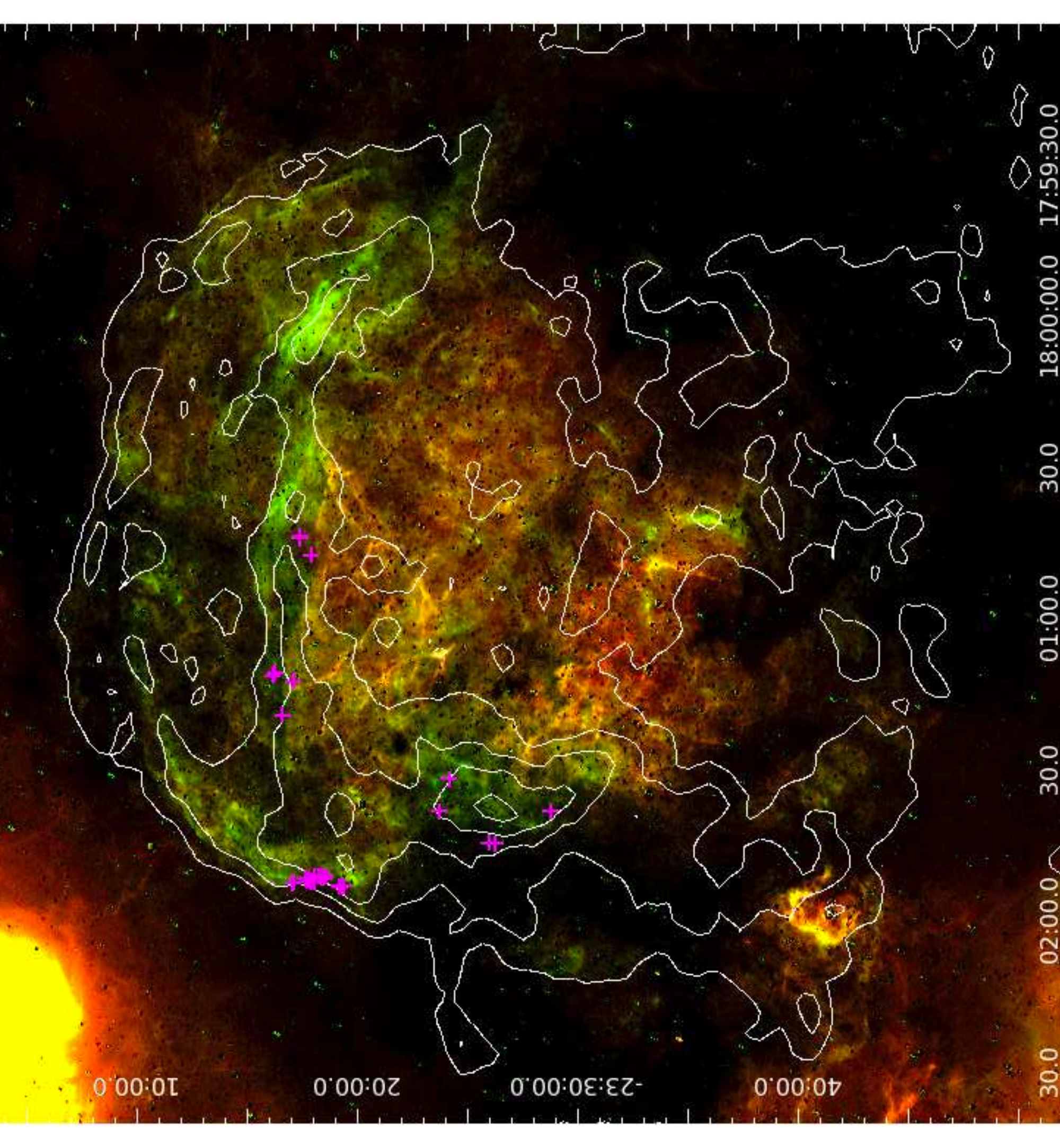}
\caption{The same color image shown in Fig.~\ref{figure_w28_color}, with contours from the 1415 
MHz radio image \citep{Dubner00} overlaid for comparison. The contour levels are 0.002,
0.012, 0.022, 0.032 and 0.042 Jy/beam. The locations of OH masers 
\citep{Claussen99} are shown as magenta crosses: these masers are associated with
interactions between SNRs and molecular clouds. 
Note that the \sii\ emission is concentrated along the eastern and northern radio
rim of the SNR, and that the masers are preferentially correlated with the regions of
\sii\ emission.} 
\label{figure_w28_color_overlay}
\end{figure} 

\clearpage
\begin{figure}
\includegraphics{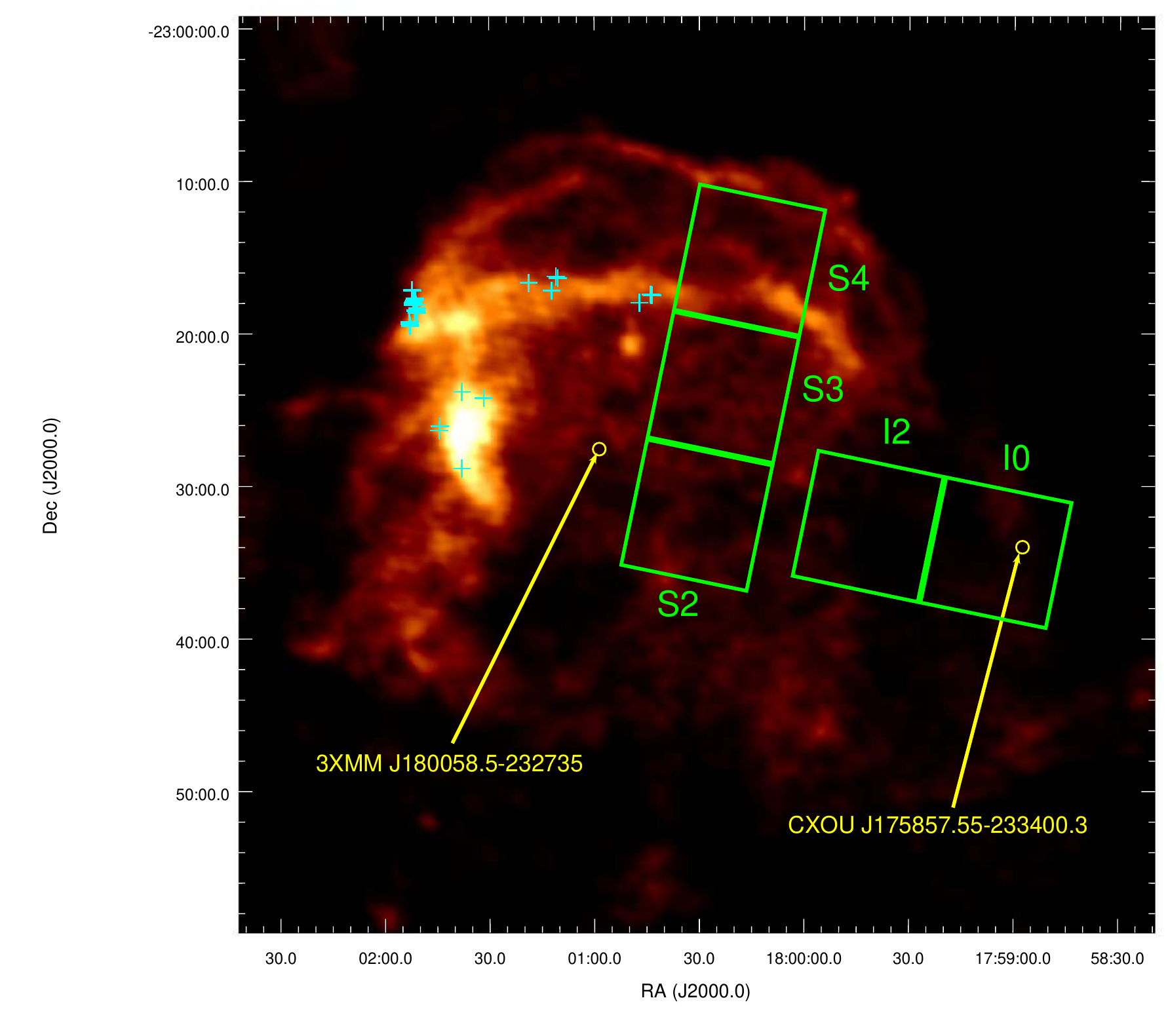}
\caption{Radio map of W28 made with the VLA \citep{Dubner00}. The green squares denote the 
footprints of the ACIS chips during the {\it Chandra} observation of this SNR.
The cyan crosses indicate the locations of the 1720 MHz OH masers detected by 
\citet{Claussen99} and the yellow circles indicate the locations of the discrete X-ray sources
3XMM J180058.5$-$232735 and CXOU J175857.55$-$233400.3 (see 
Section \ref{DiscreteXraySourceSection}).}
\label{W28VLARadioMasers}
\end{figure}

\clearpage
\begin{figure}
\includegraphics{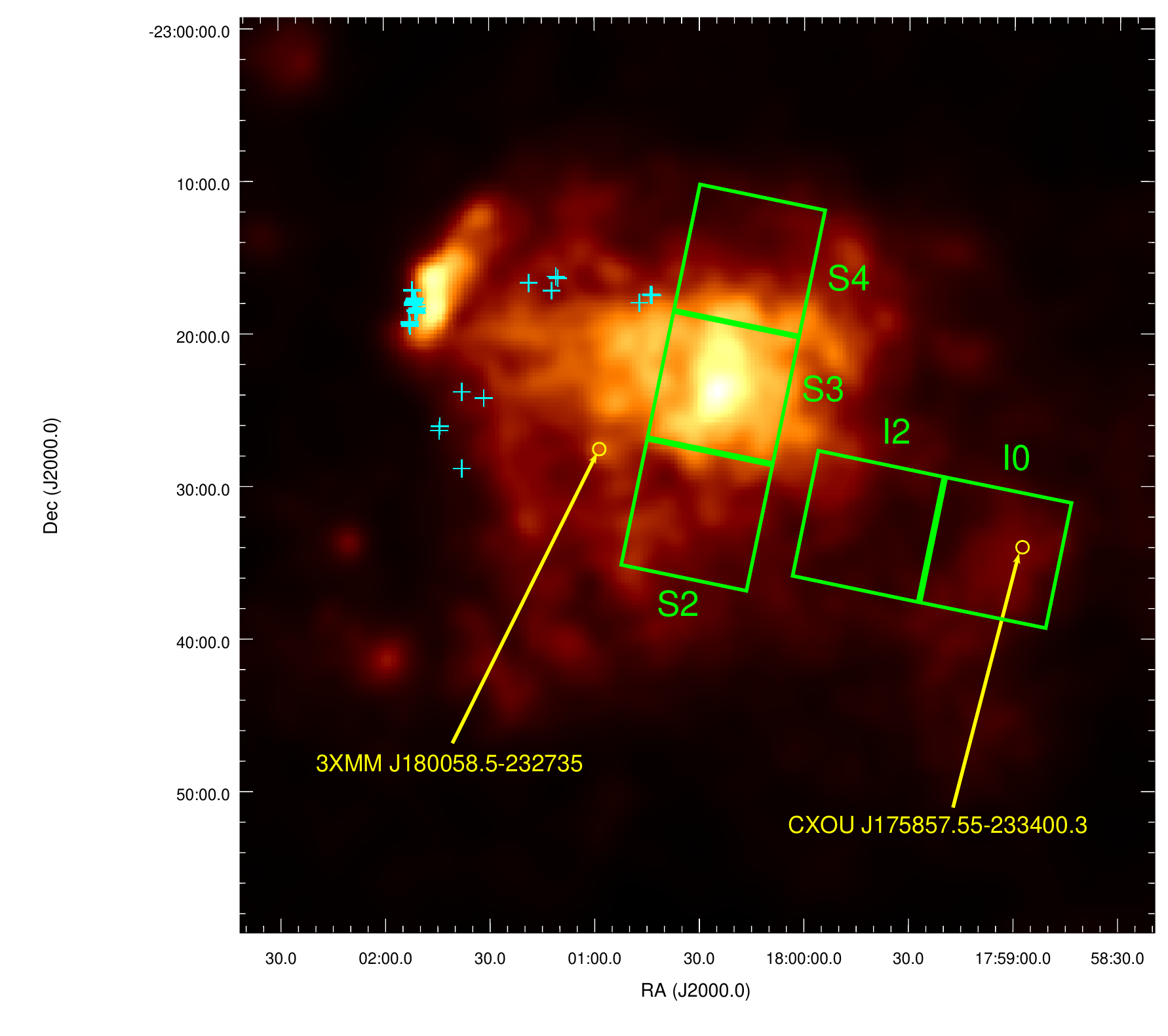}
\caption{{\it ROSAT} PSPC image of W28 \citep{Rho02}. Similar to Figure \ref{W28VLARadioMasers}, the
locations of the footprints of the ACIS chips, the OH masers and the locations of discrete X-ray sources
discussed in Section \ref{DiscreteXraySourceSection} are indicated. Notice the centrally-concentrated X-ray
morphology of W28 in contrast to the shell-like radio morphology depicted in Figure \ref{W28VLARadioMasers}:
these contrasting morphologies characterize MMSNRs.}
\label{W28ROSATPSPC}
\end{figure}

\clearpage
\begin{figure}
\includegraphics{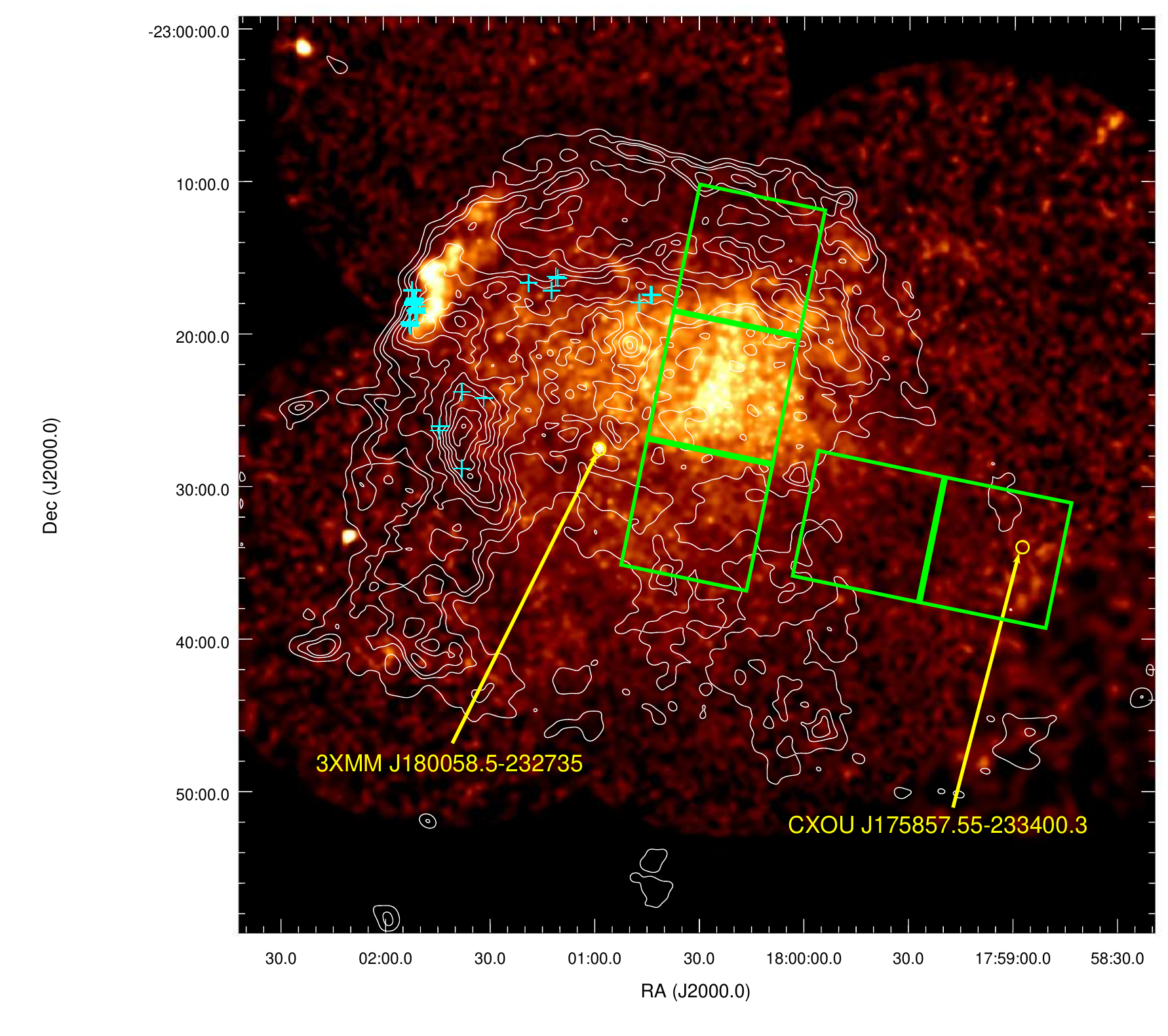}
\caption{Adaptively-smoothed (with a smoothing kernel of 30 counts)
{\it ROSAT} HRI mosaic of W28 as prepared
from the set of HRI observations listed in Table \ref{Table1}. Radio emission from W28 is depicted using
overlaid white contours: these contours have been placed at the levels of 0.0035, 0.0075, 0.01, 
0.015, 0.02, 0.025, 0.03, 0.035, 0.041, 0.0475, 0.05375 and 0.06 Jansky/beam. Similar to Figure 
\ref{W28VLARadioMasers}, the
locations of the footprints of the ACIS chips, the OH masers and the locations of discrete X-ray sources
discussed in Section \ref{DiscreteXraySourceSection} are indicated. The contrasting X-ray and radio
morphologies of W28 are again apparent in this figure.} \label{W28ROSATHRIMasers}
\end{figure}

\clearpage
\begin{figure}
\plotone{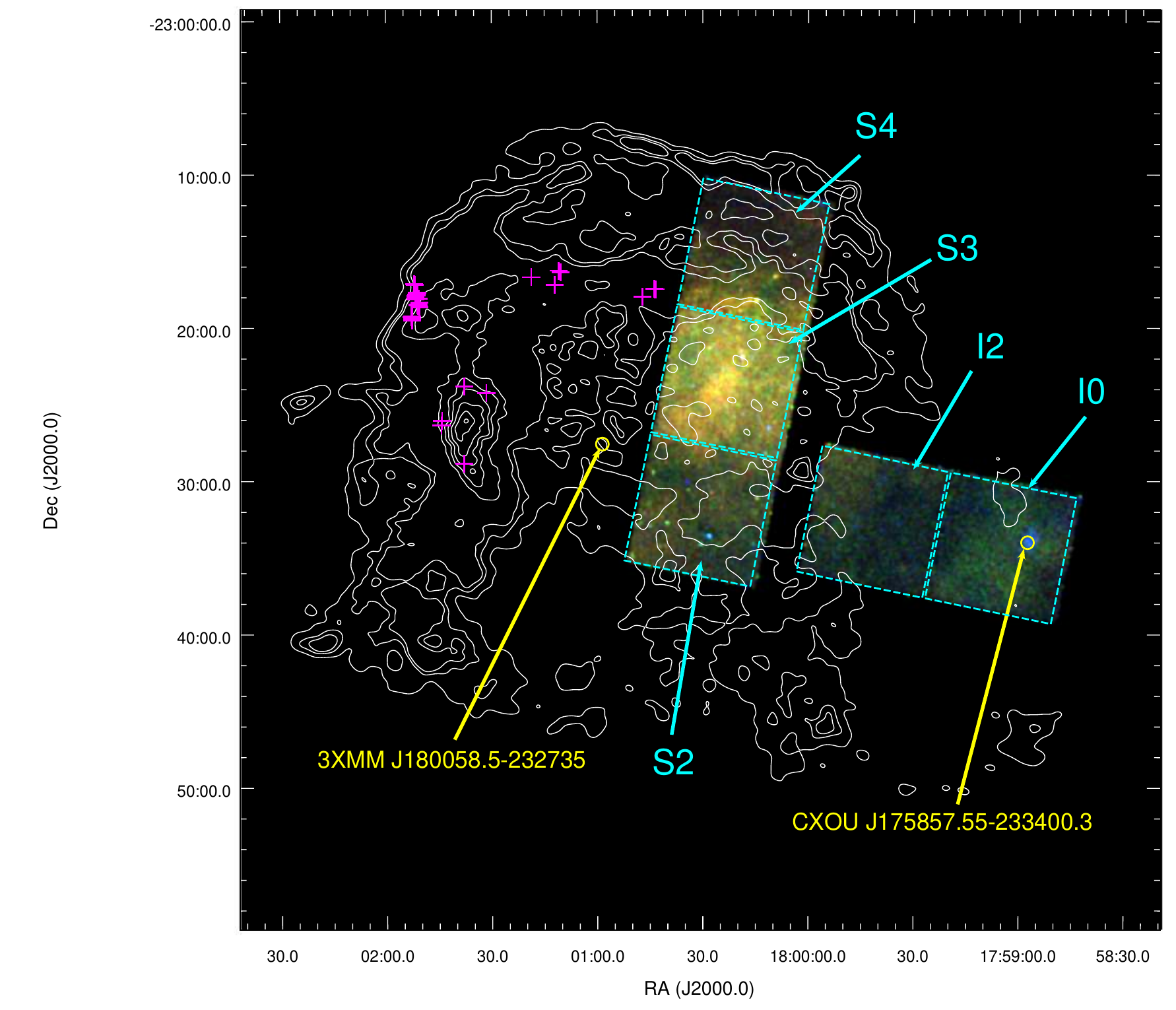}
\caption{Adaptively smoothed {\it Chandra} ACIS image of W28. Soft (0.5 keV - 1.2 keV), 
medium (1.2 keV - 2.0 keV) and hard (2.0 keV - 7.0 keV) emission is shown in red, green and
blue, respectively. The red contours correspond to radio emission detected by the VLA at a 
frequency of 1415 MHz: the contour levels are 0.0035, 0.0075, 0.01, 0.023, 0.03, 0.035, 
0.041, 0.0475, 0.5375 and 0.06 Jansky/beam. The green crosses show the locations of the 1720 
MHz OH masers detected by \citet{Claussen99}. \label{W28Chandra3Color}}
\end{figure}

\clearpage
\begin{figure}
\includegraphics[scale=0.65,angle=-90]{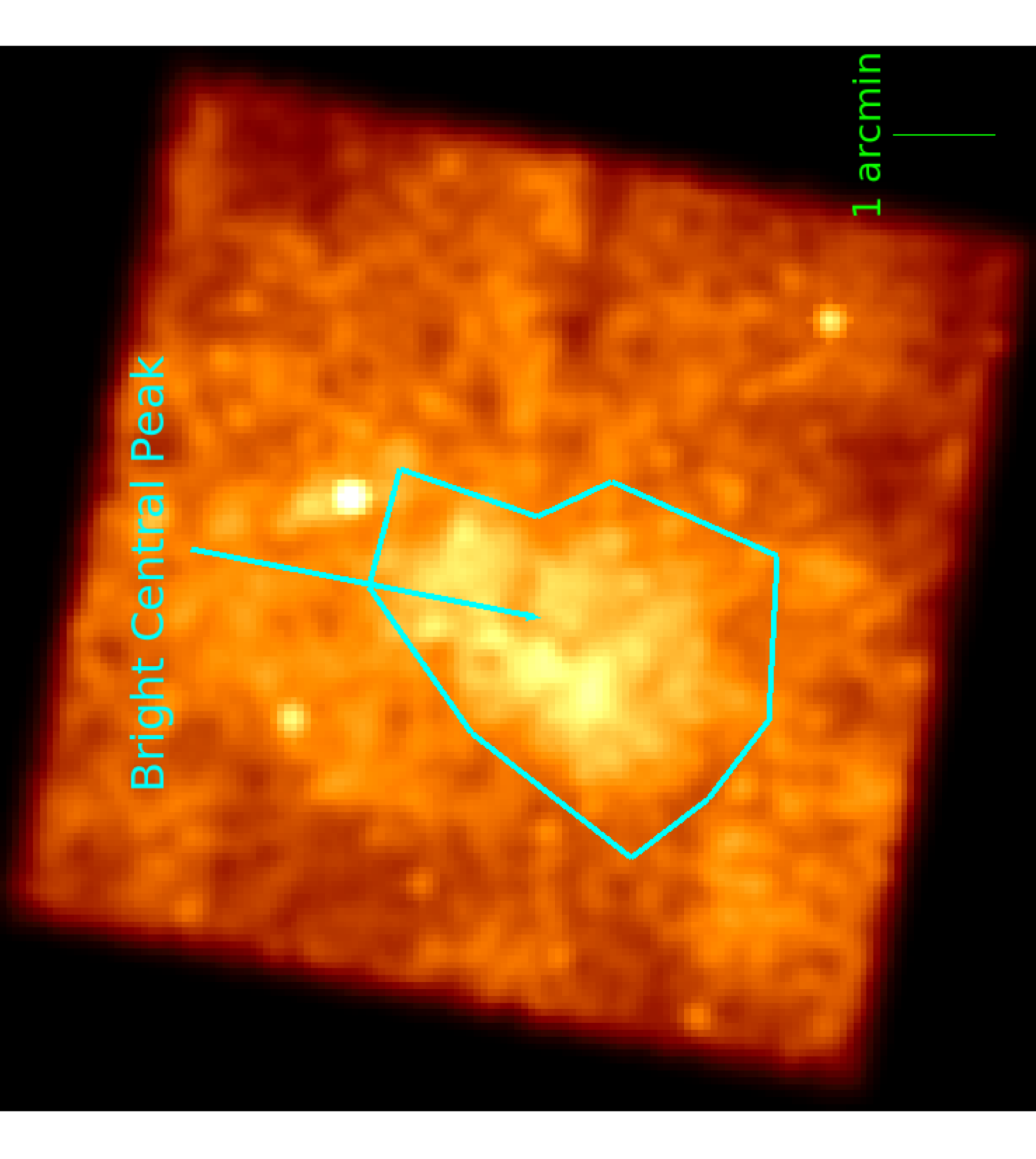}
\caption{{\it Chandra} ACIS-S3 broadband (0.5 keV - 7.0 keV) exposure-corrected and
smoothed image of the central portion of W28. The region
of spectral extraction for the bright central peak is indicated. 
\label{W28BrightCentralPeakFigure}}
\end{figure}

\clearpage
\begin{figure}
\includegraphics[scale=0.35,angle=90]{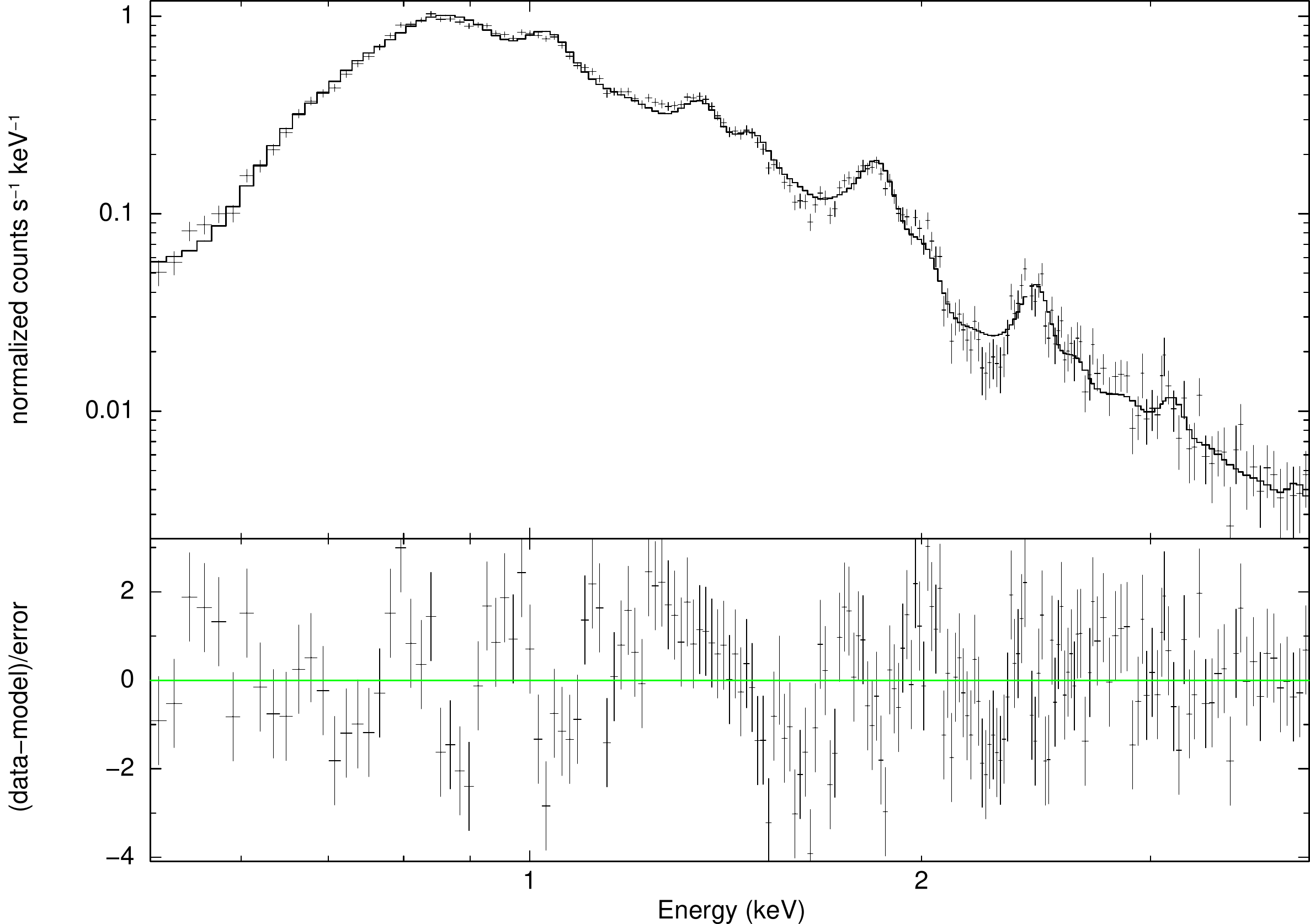}
\includegraphics[scale=0.35,angle=90]{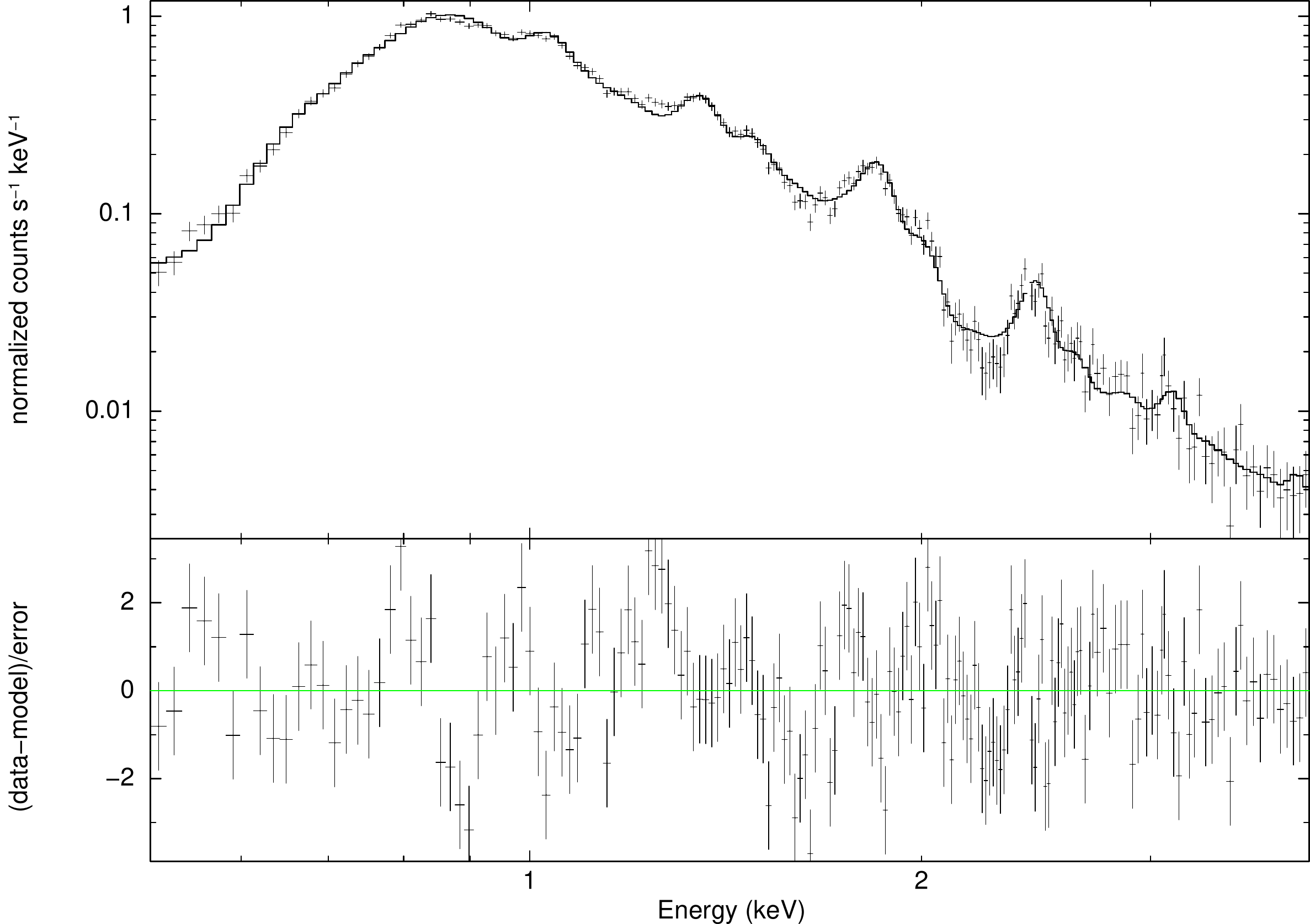}
\includegraphics[scale=0.35,angle=90]{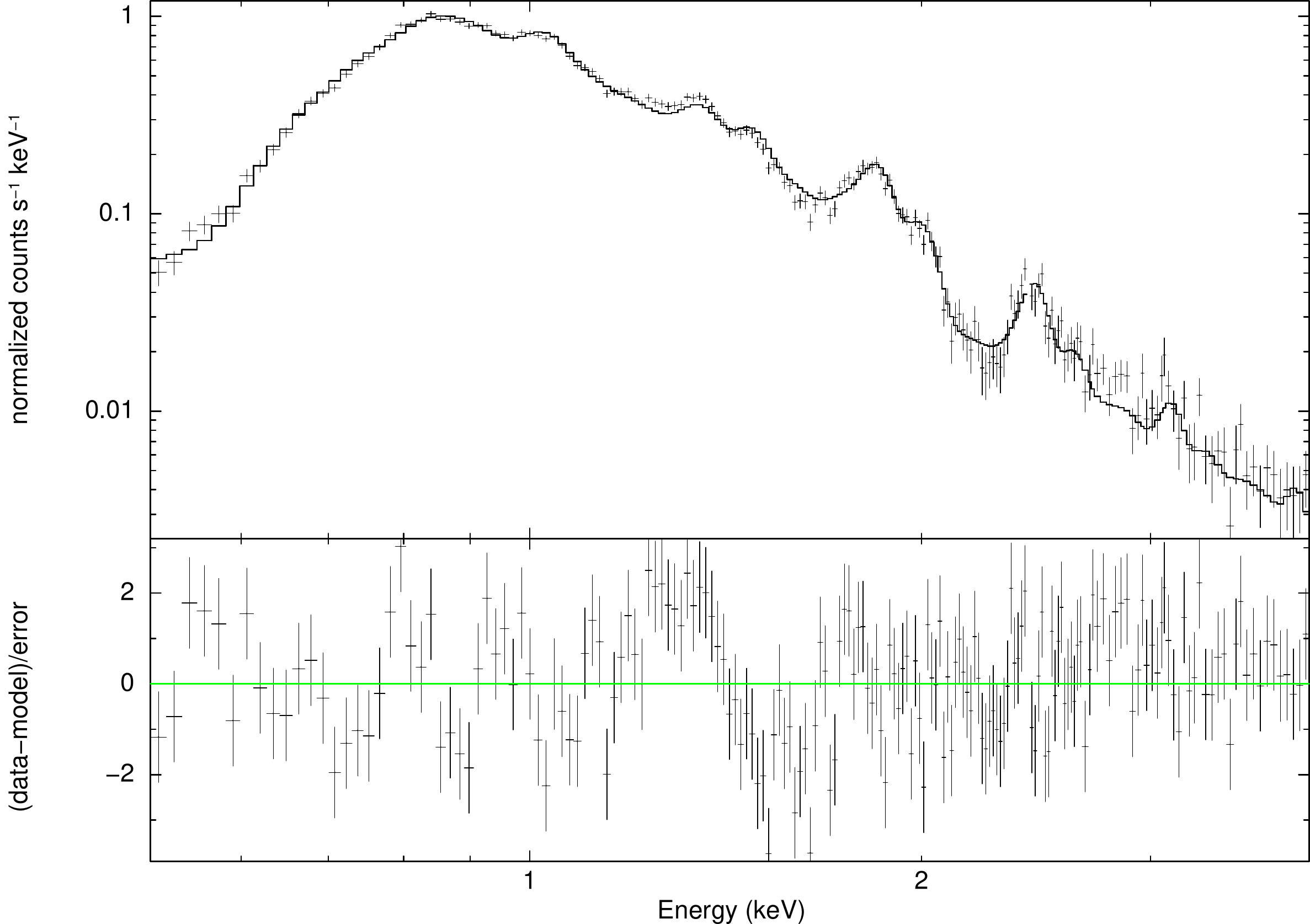}
\includegraphics[scale=0.35,angle=90]{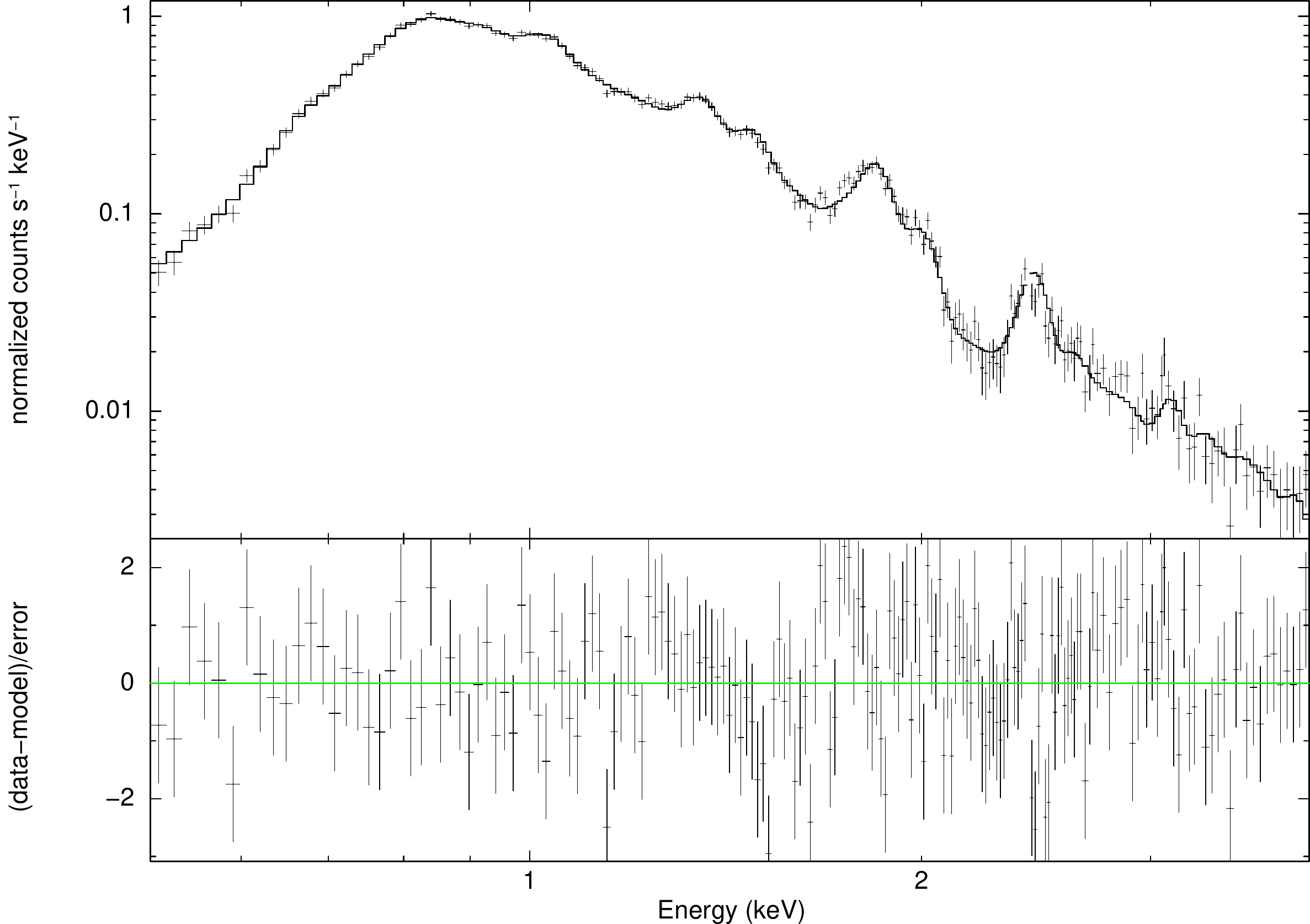}
\caption{Four spectral fits to the extracted spectrum of the central bright emission region of W28.
The parameters of these fits are listed in Table \ref{BlobSpectrumTable}. Upper left:
TBABS$\times$(VAPEC+VAPEC) fit. Upper right: TBABS$\times$(VNEI+VNEI) fit. Lower
left: TBABS$\times$VRNEI. Lower right: TBABS$\times$(VRNEI+VRNEI).}
\label{CentralRegionSpectralFigure}
\end{figure}

\clearpage
\begin{figure}
\includegraphics[scale=1.00]{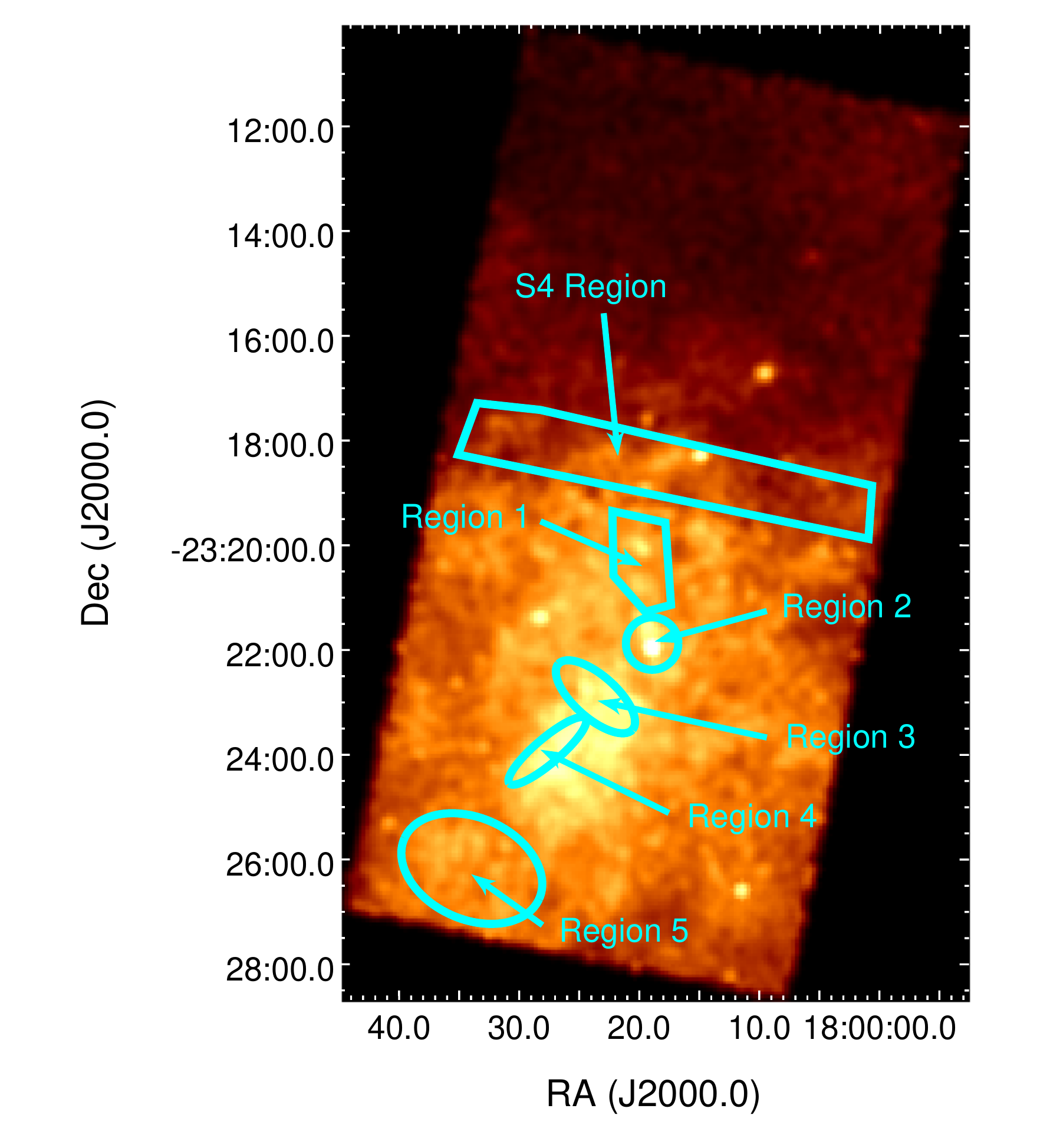}
\caption{An exposure-corrected image of the ACIS-S3 and ACIS-S4 chips over the broad
energy range (0.7 - 5.0 keV). The five regions of spectral extraction on the ACIS-S3 chip
and the region of spectral extraction on the ACIS-S4 chip are indicated.}
\label{ChandraSpecExtractionFigure}
\end{figure}

\clearpage
\begin{figure}
\includegraphics[scale=0.35,angle=90]{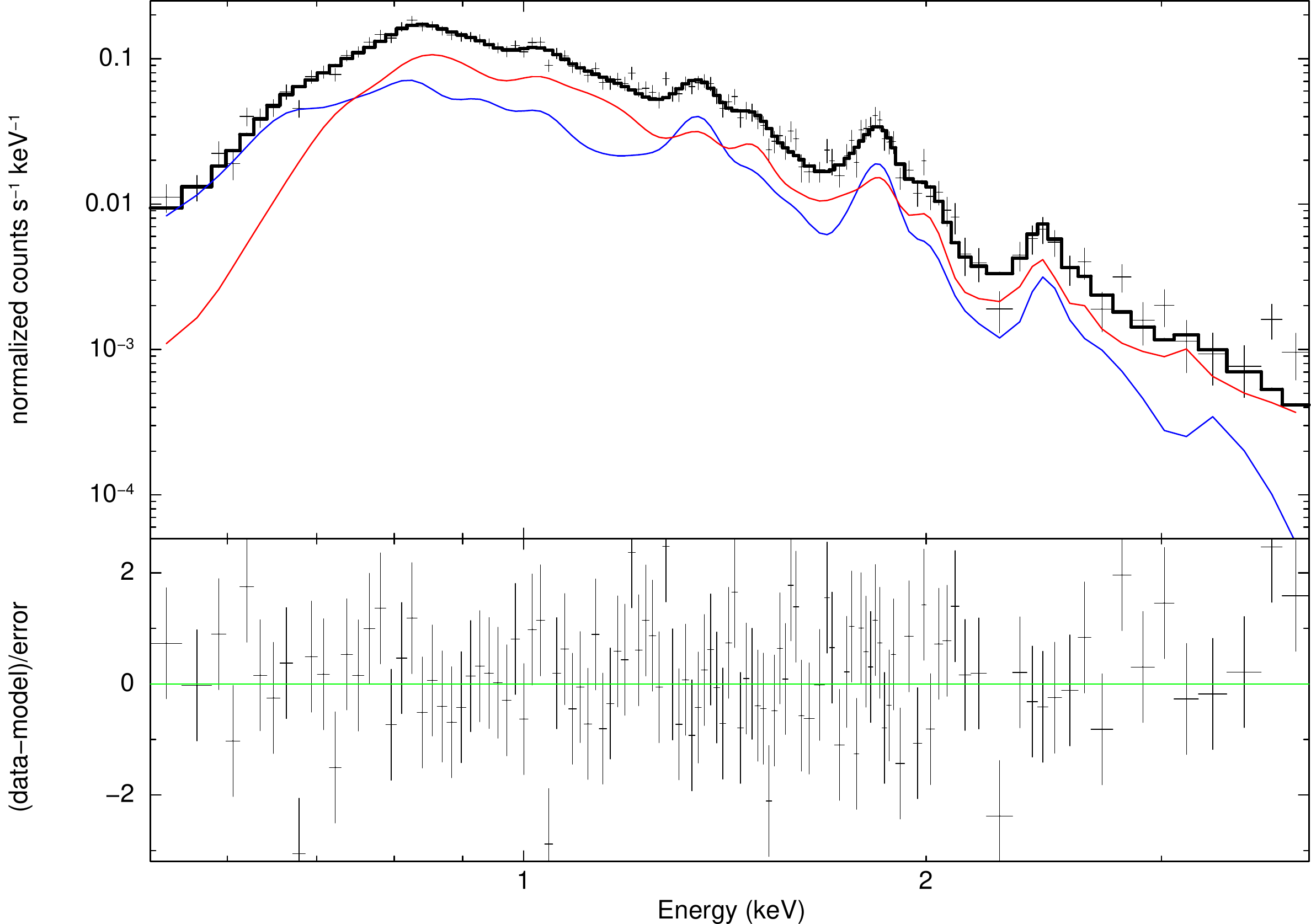}
\includegraphics[scale=0.35,angle=90]{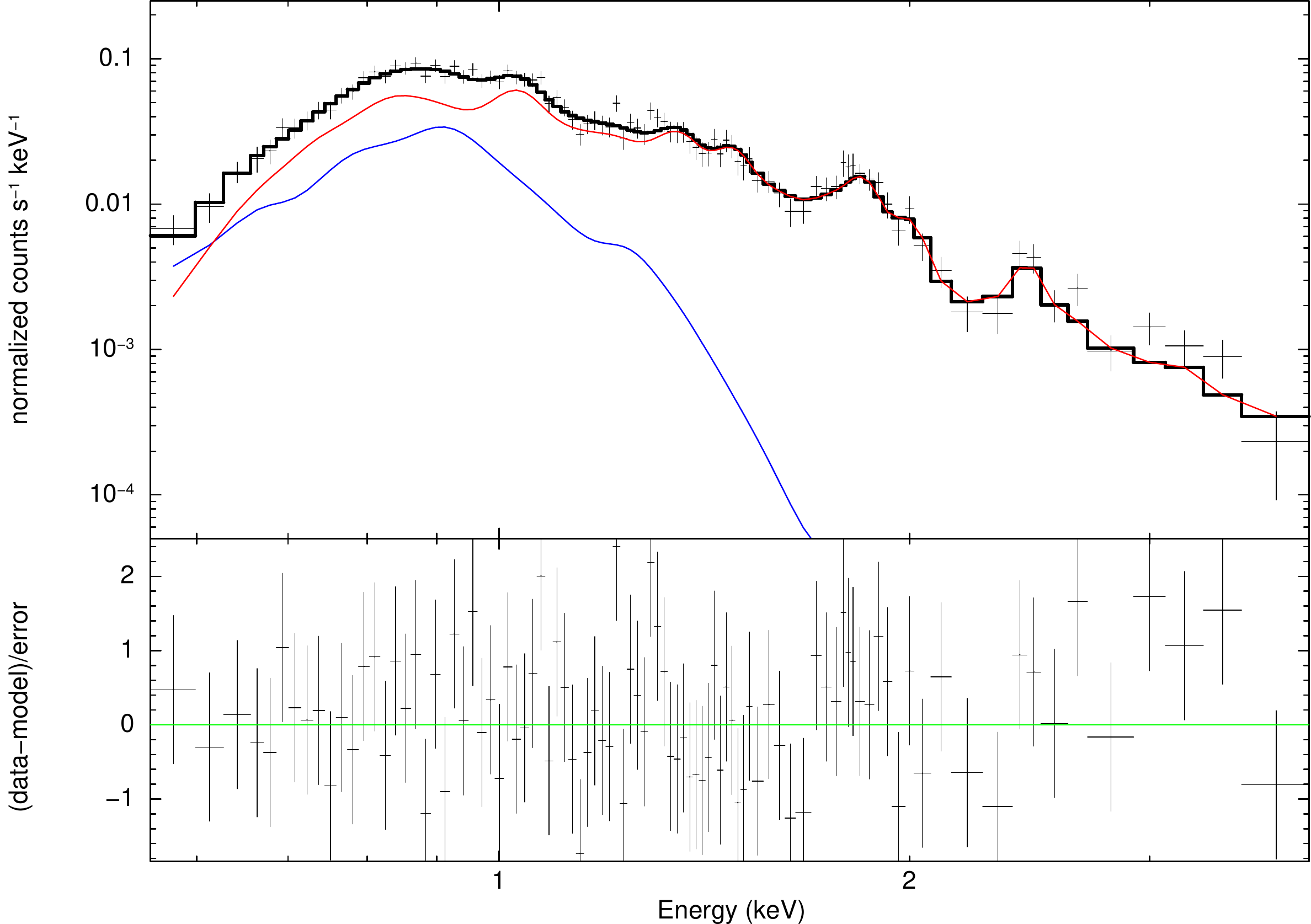}
\includegraphics[scale=0.35,angle=90]{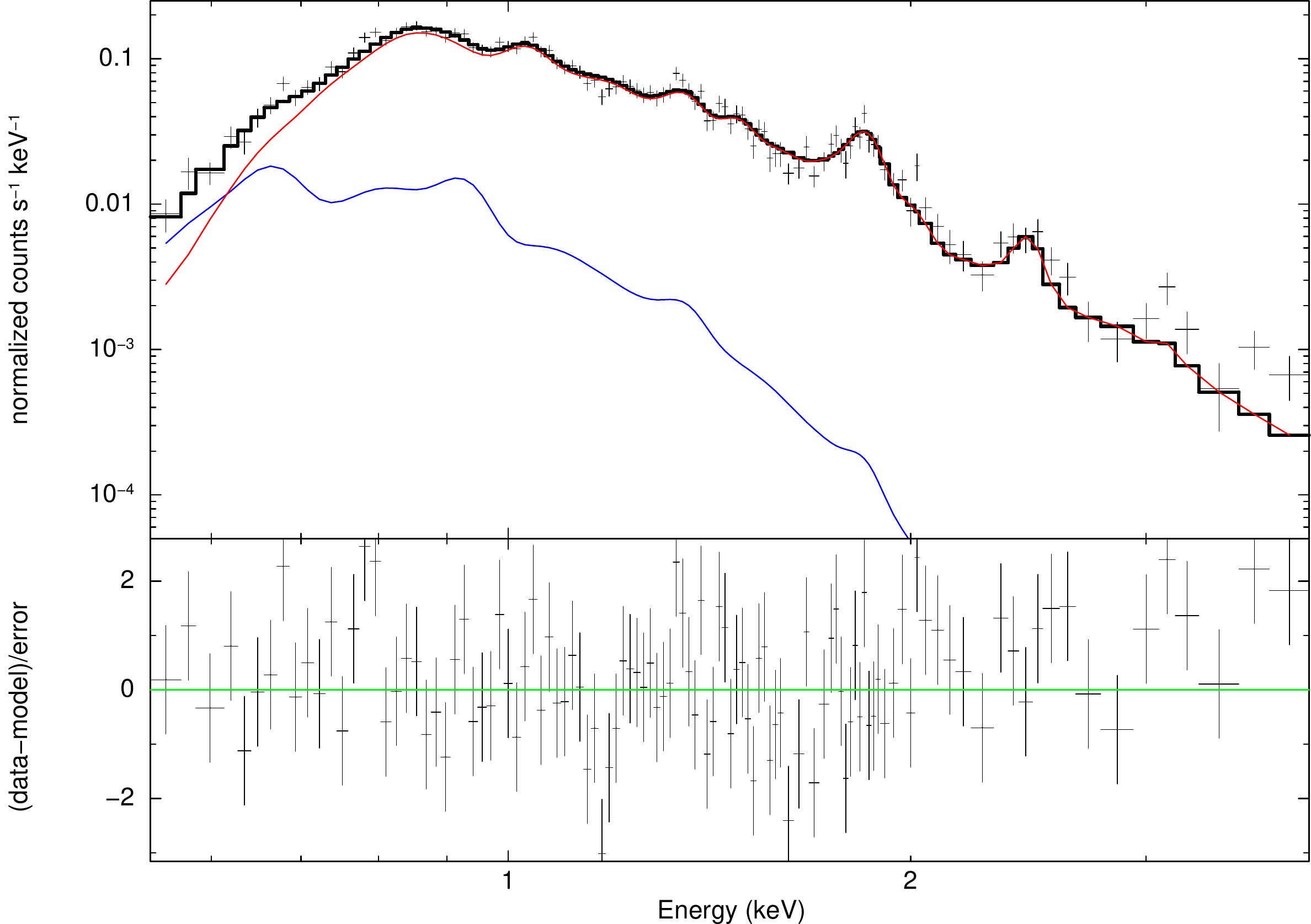}
\includegraphics[scale=0.35,angle=90]{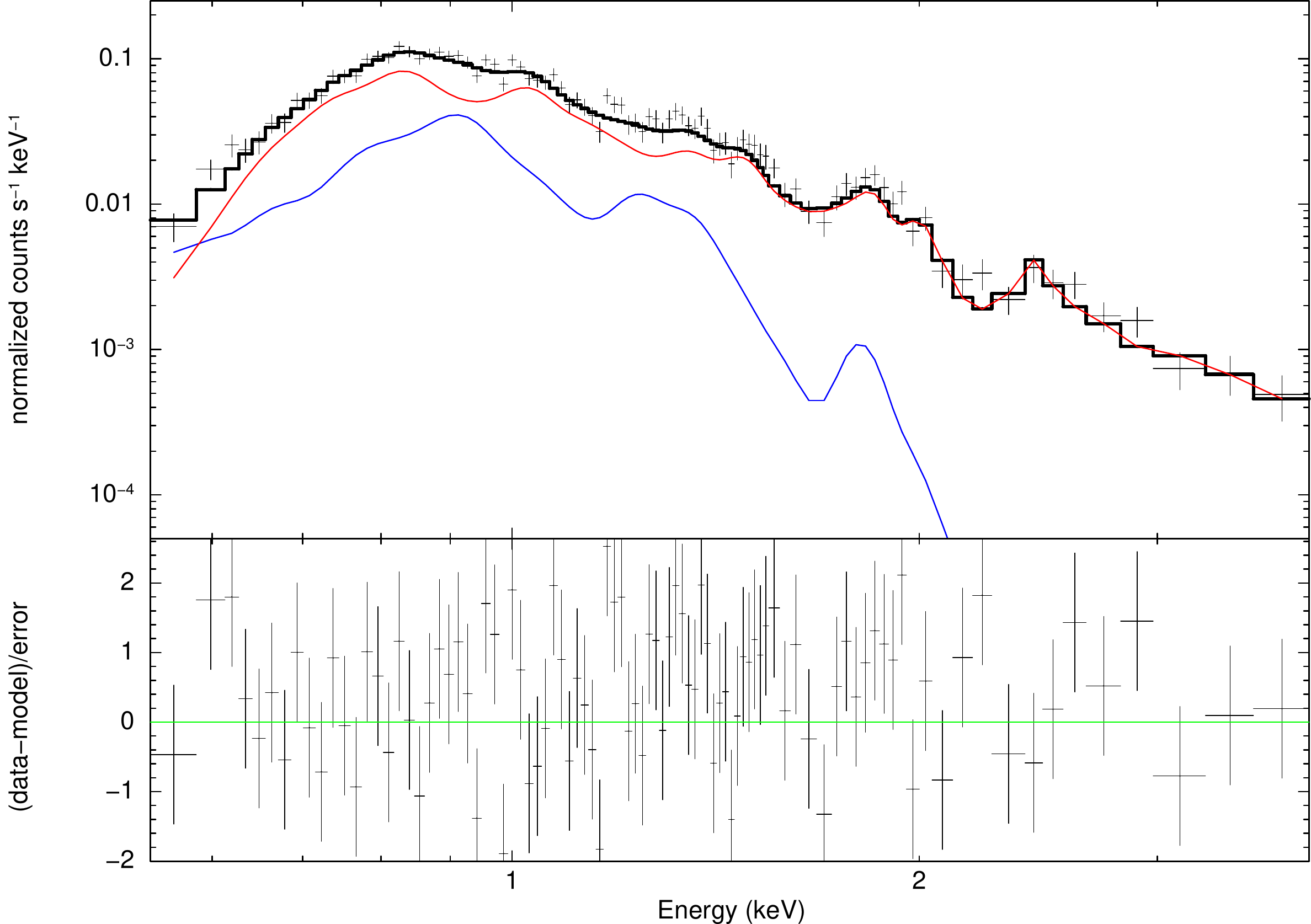}
\includegraphics[scale=0.35,angle=90]{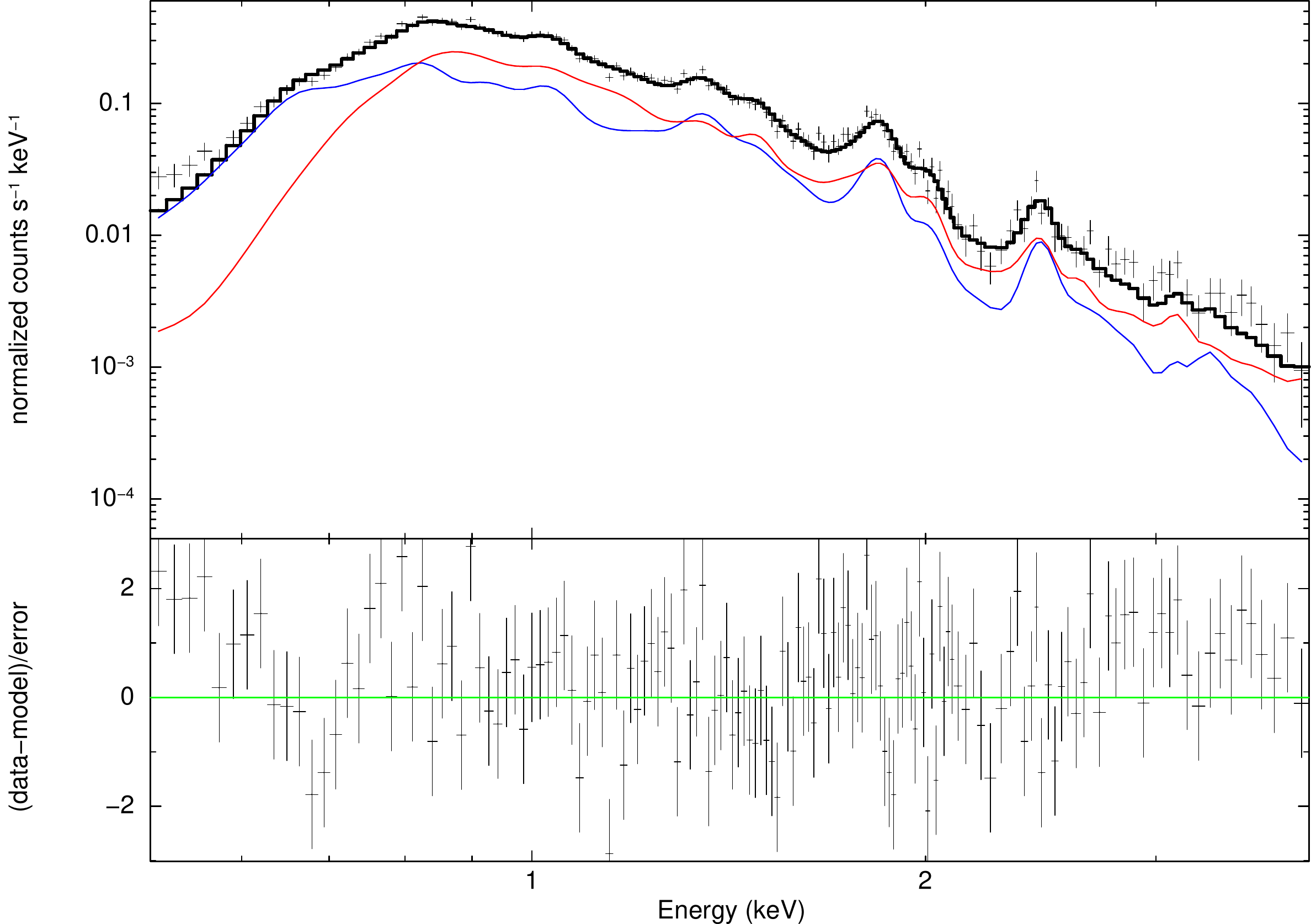}
\includegraphics[scale=0.35,angle=90]{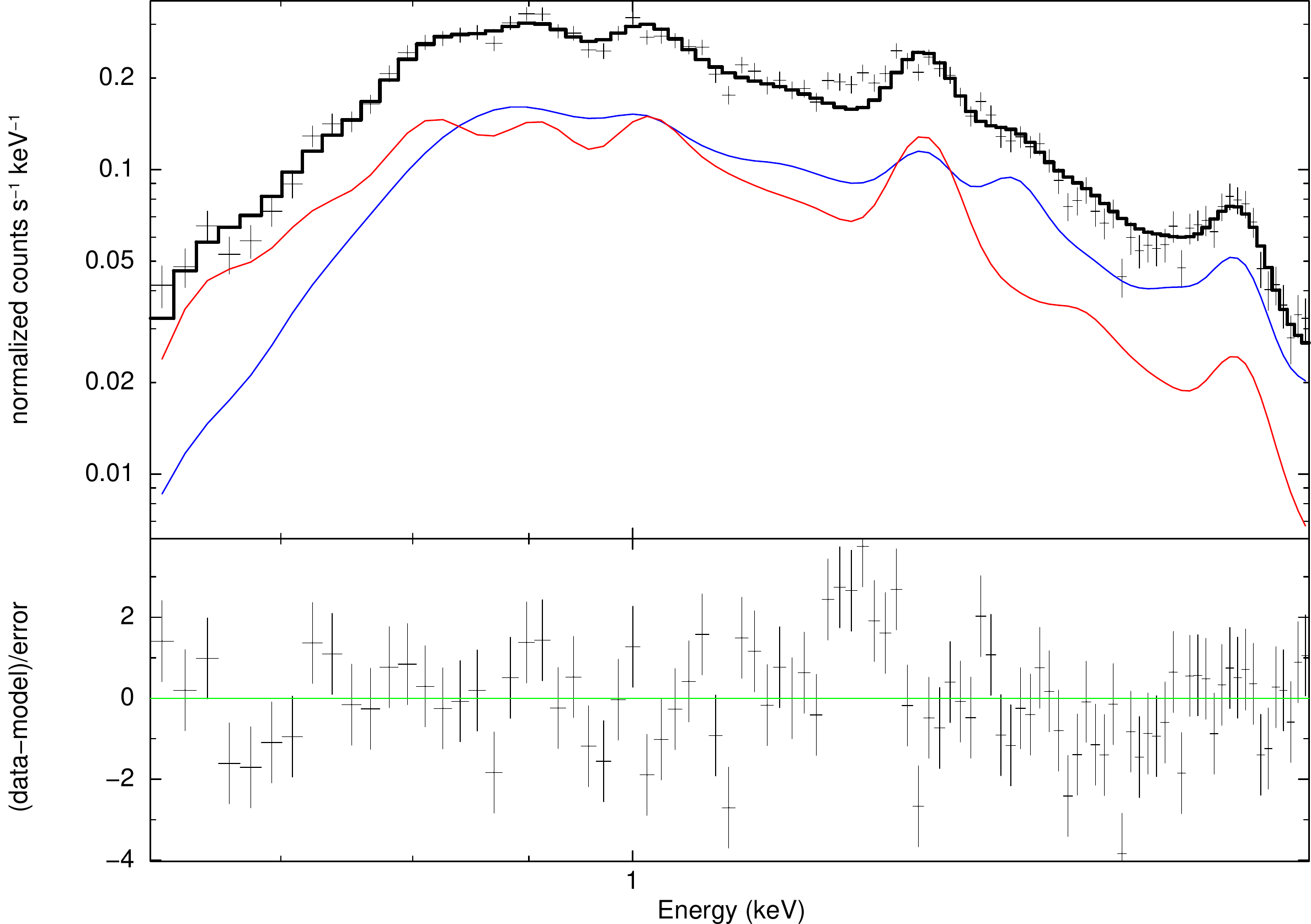}
\caption{\small{Fits to the extracted spectra of the five ACIS-S3 regions and the ACIS-S4
region using a TBABS$\times$(VRNEI+VRNEI) model: the parameters of the fits to these 
models are given in Table \ref{SixRegionsVRNEIVRNEIFitsTable}. In these images, the 
emission from the $k$$T$$_{\rm{e1}}$ and 
$k$$T$$_{\rm{Z1}}$ component is shown in blue while the emission from the $k$$T$$_{\rm{e2}}$ and 
$k$$T$$_{\rm{Z2}}$ component is shown in red. Upper left: Region 1. Upper right: Region 2. 
Middle left: Region 3. Medium right: Region 4. Bottom left: Region 5. Bottom right: ACIS-S4
region.}}
\label{SixRegionsSpectraFigure}
\end{figure}

\clearpage
\begin{figure}
\includegraphics[scale=0.65,angle=-90]{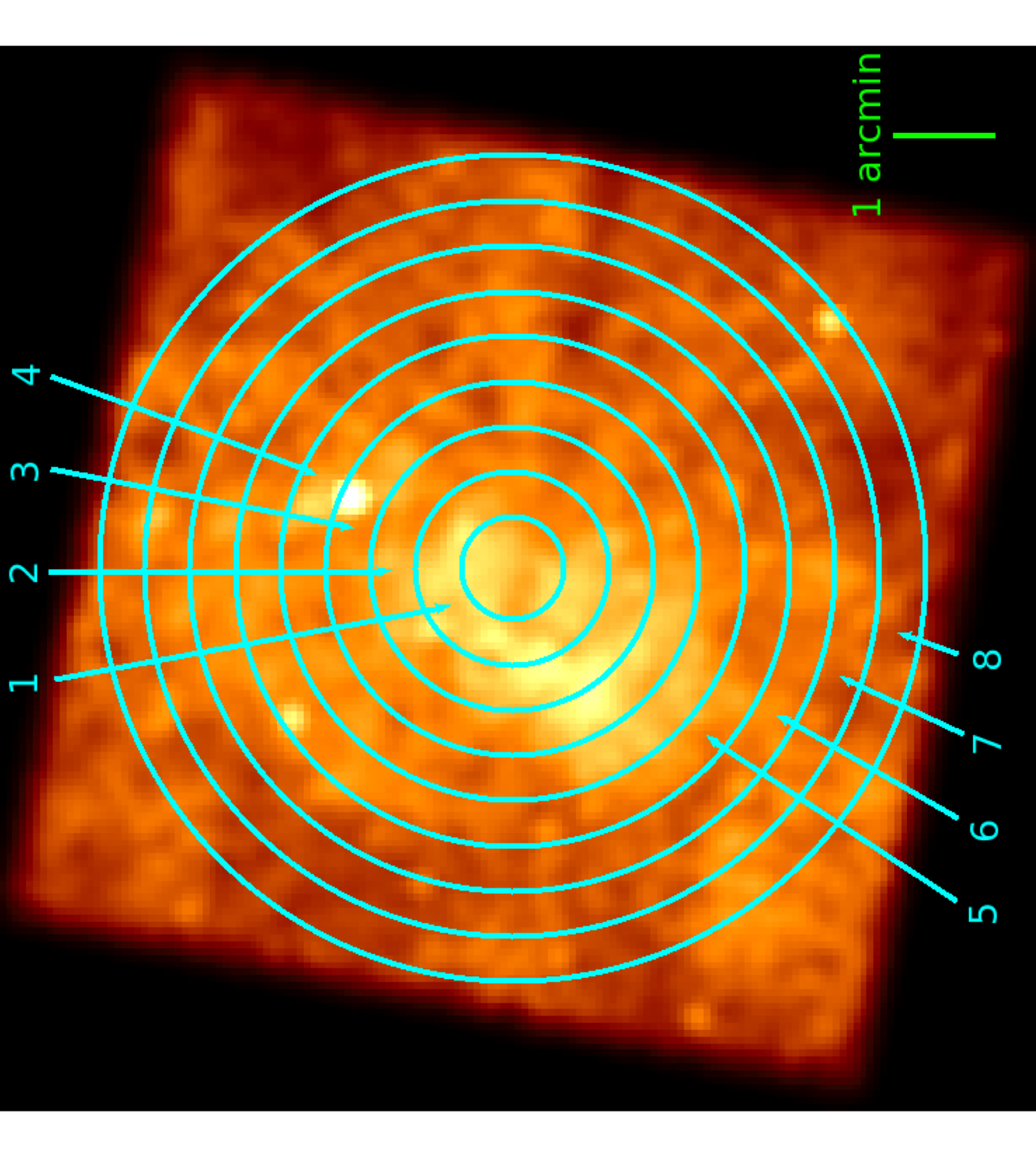}
\caption{Same as Figure \ref{W28BrightCentralPeakFigure} but with the annular regions of
spectral extraction indicated.}
\label{AnnularRegionsFigure}
\end{figure}

\clearpage
\begin{figure}
\includegraphics[scale=0.28]{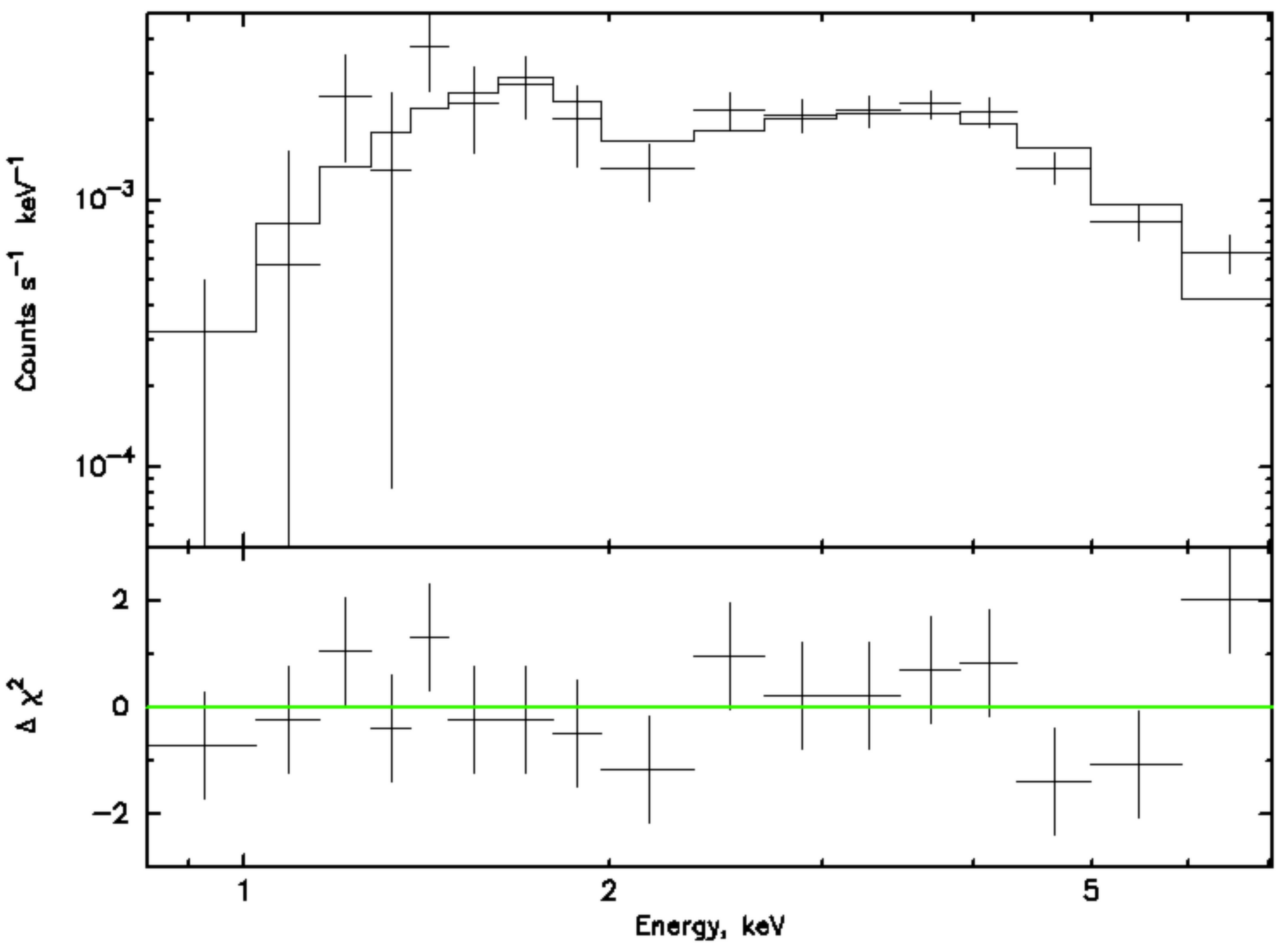}
\includegraphics[scale=0.28]{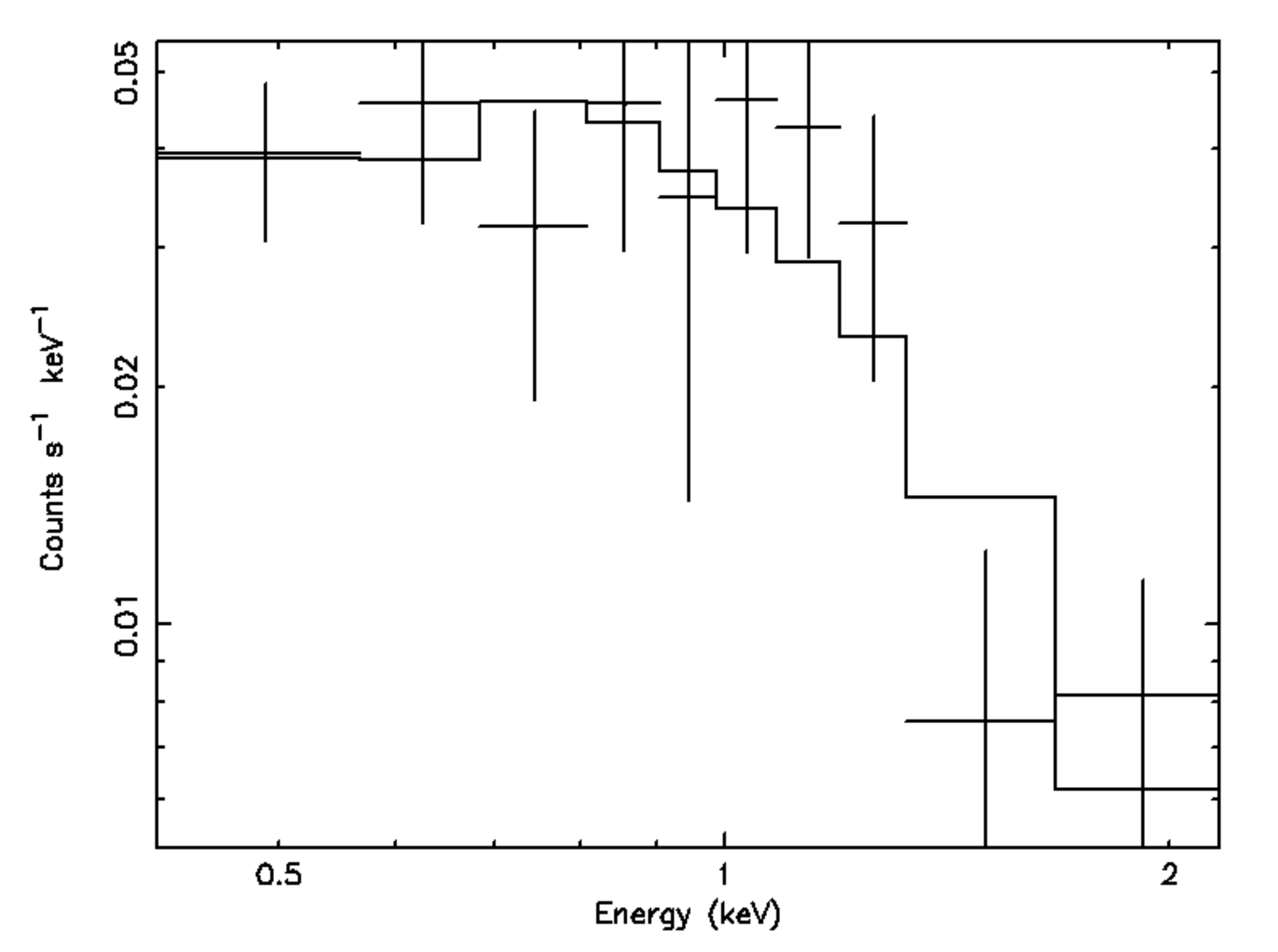}
\caption{The ACIS spectrum of CXOU J175857.55$-$233400.3 (left) and the EPIC-PN spectrum of 
3XMM J180058.5$-$232735 (right). Both spectra are fitted with an absorbed power law model 
(see Section \ref{DiscreteXraySourceSection}).}
\vspace{-0.0cm}
\label{PointSourceSpectra}
\end{figure}

\begin{figure*}
\includegraphics[scale=0.5,trim=0 0 0 0]{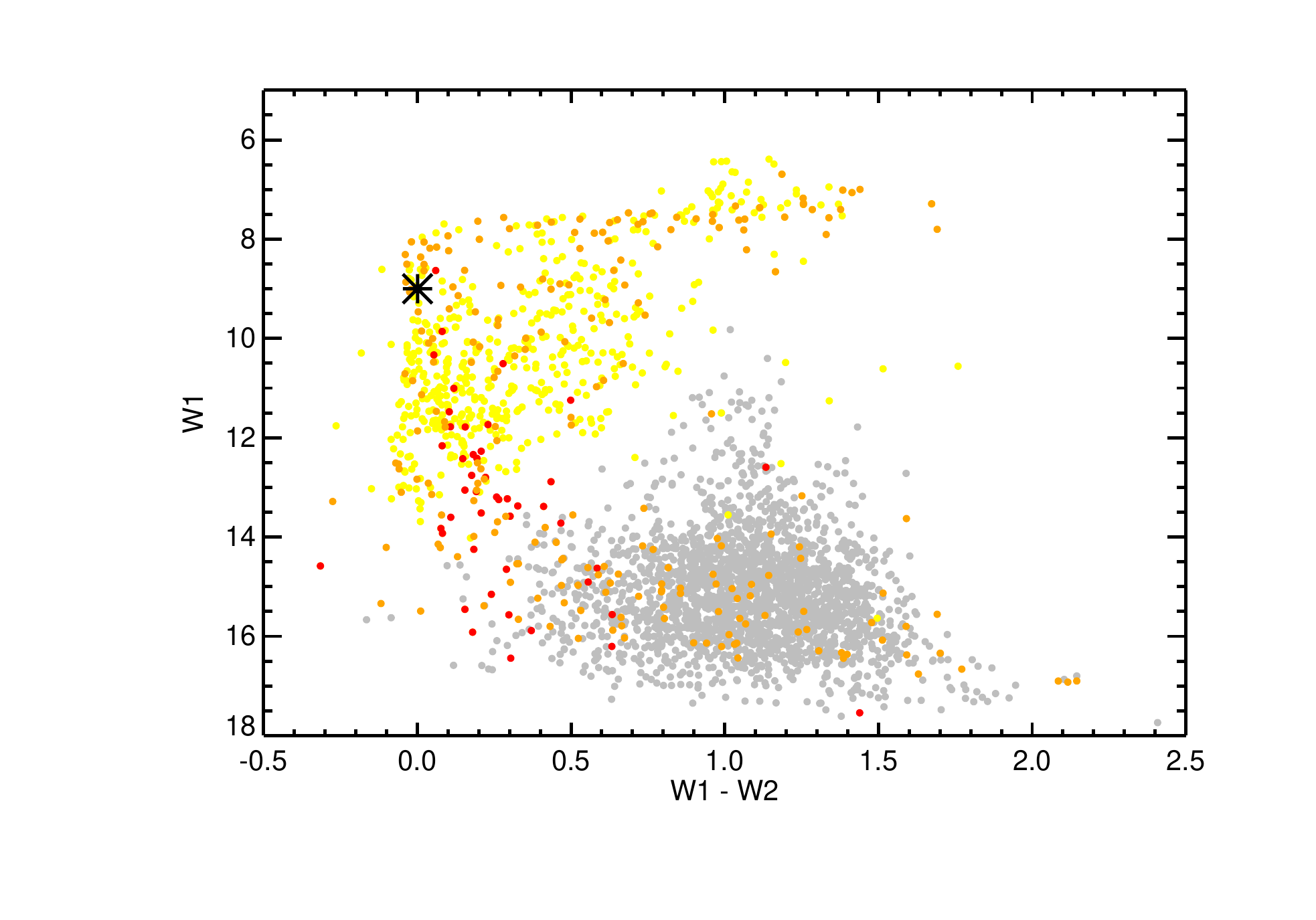}
\includegraphics[scale=0.5,trim=80 0 0 0]{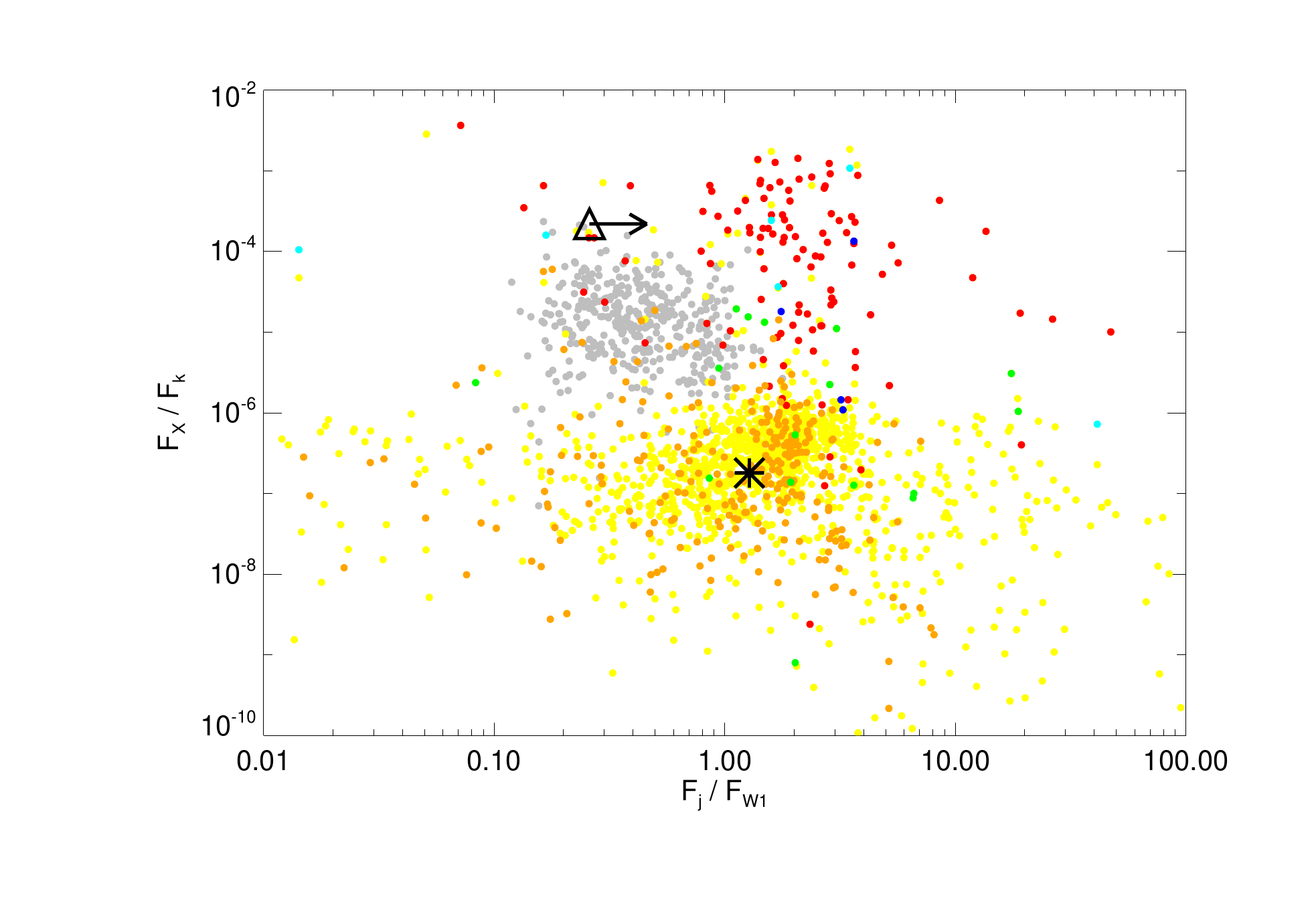}
\includegraphics[scale=0.5,trim=0 50 0 60]{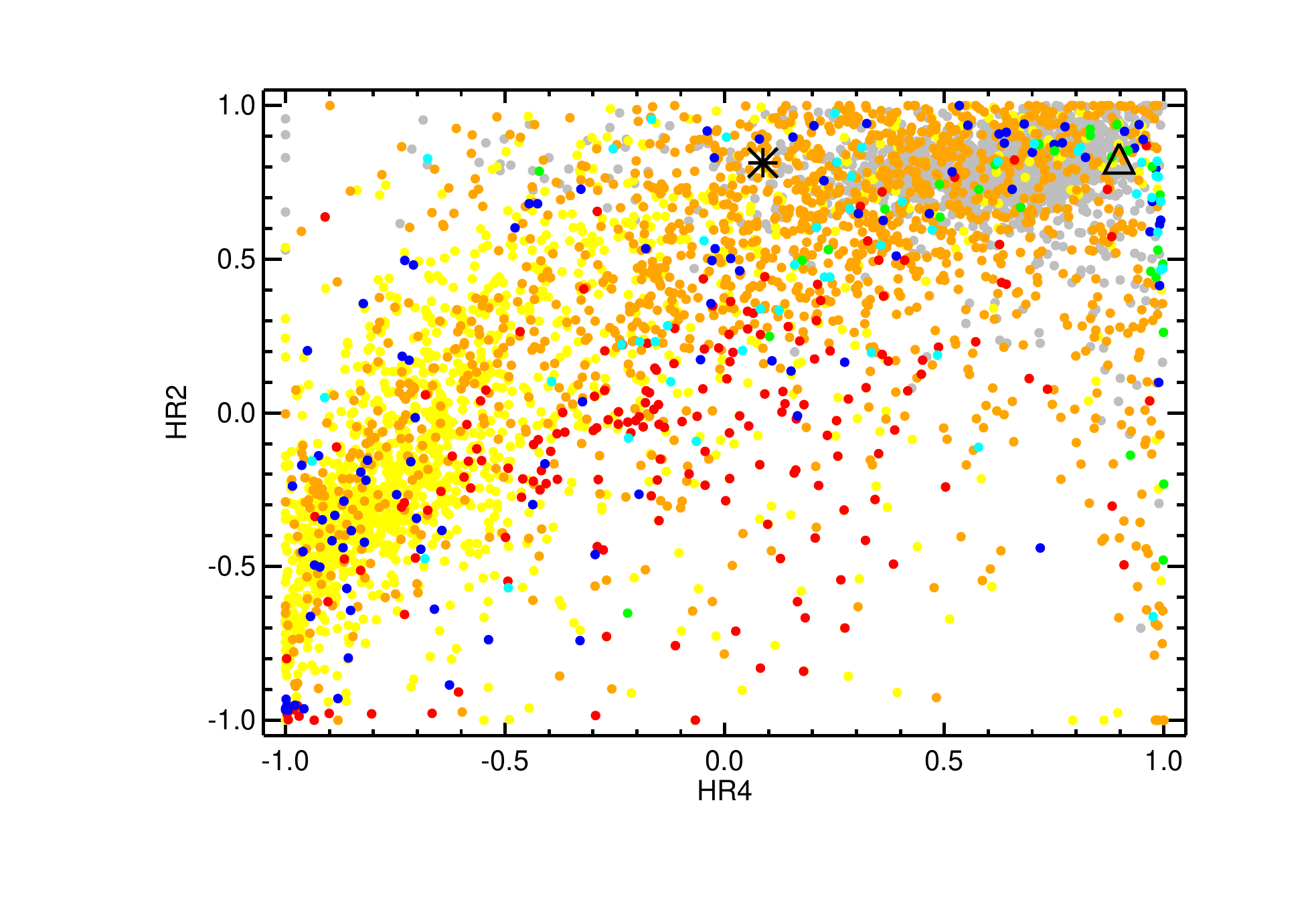}
\caption{$\it{WISE}$ color-color diagram (upper left), $F$$_{j}$/$F$$_{W1}$ vs. 
$F$$_{X}$/$F$$_{k}$ flux diagram (upper right) and HR4 vs. HR2 hardness ratio diagram
(lower left) diagrams. The plotted black triangles and asterisks corresponds to the locations of
of CXOU J175857.55$-$233400.3 and 2XMM J180058.6$-$232724 = 3XMM J180058.5$-$232735,
respectively. The arrow indicates a lower limit. Other symbols are AGN (gray), Stars 
(yellow), YSOs (orange), CVs (red), pulsars (blue), HMXBs (green) and LMXBs (cyan) based on the 
training dataset described in \citet{Sonbas16}.}
\label{ColorColorFigures}
\end{figure*}

\begin{figure*}
\includegraphics[scale=0.6,angle=-90]{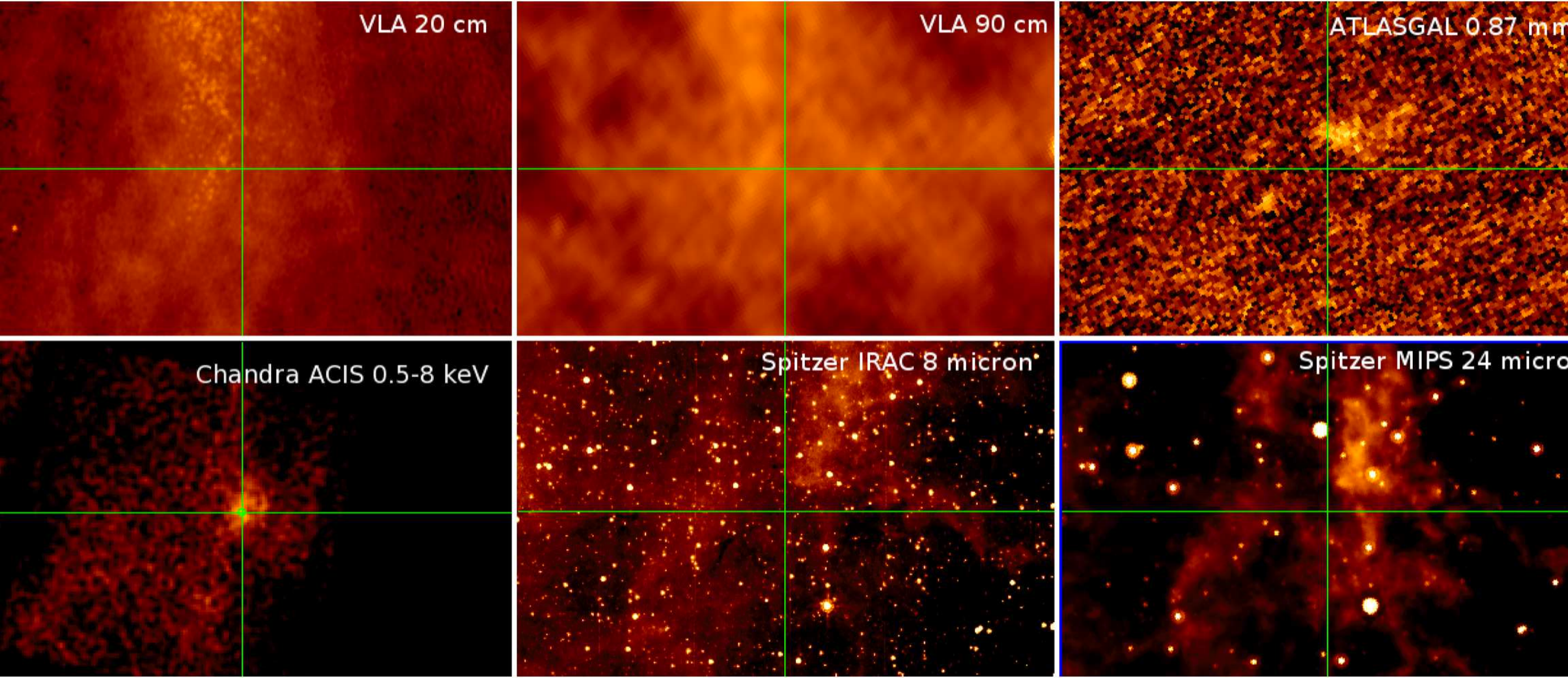}
\caption{The vicinity of the CXOU J175857.55$-$233400.3 at different wavelengths. The source 
position is indicated by crosshairs in all panels.}
\label{HardSourceMultiWavelengthFigures}
\end{figure*}

\end{document}